\documentclass[twocolumn]{aastex63}

\usepackage{savesym}
\savesymbol{tablenum}
\usepackage{siunitx}
\restoresymbol{SIX}{tablenum}

\usepackage{natbib}
\usepackage{array}
\usepackage{rotating, booktabs}  
\usepackage{graphicx}
\usepackage[tbtags]{amsmath}
\usepackage{amssymb}
\usepackage{amsthm}

\usepackage{bm}
\usepackage{mathrsfs}
\usepackage{mathtools}
\usepackage{empheq, nccmath}
\usepackage{aas_macros}
\usepackage{xspace}
\usepackage{layouts}
\usepackage{makecell}

\makeatletter
\AtBeginDocument{\let\LS@rot\@undefined}
\makeatother

\usepackage{pdflscape} 

\usepackage{lineno}

\newcommand{\Abacus}{\textsc{Abacus}\xspace}
\newcommand{\AbacusSummit}{\textsc{AbacusSummit}\xspace}

\newcommand{\Compaso}{\textsc{CompaSO}\xspace}

\def\code#1{\texttt{#1}}

\newcommand{\hGpc}{\ensuremath{{h^{-1}\rm\,Gpc}}}
\newcommand{\Mpc}{\ensuremath{\rm\,Mpc}}
\newcommand{\hMpc}{\ensuremath{{h^{-1}\rm\,Mpc}}}

\newcommand{\hGpcC}{\ensuremath{{h^{-3}{\rm\,Gpc}^3}}}

\newcommand{\hMsun}{\ensuremath{h^{-1}{\rm\, M}_\odot}}

\makeatletter
\newcommand*\bigcdot{\mathpalette\bigcdot@{.5}}
\newcommand*\bigcdot@[2]{\mathbin{\vcenter{\hbox{\scalebox{#2}{$\m@th#1\bullet$}}}}}
\makeatother
\usepackage{scalerel,stackengine}
\stackMath
\newcommand\reallywidehat[1]{%
\savestack{\tmpbox}{\stretchto{%
  \scaleto{%
    \scalerel*[\widthof{\ensuremath{#1}}]{\kern-.6pt\bigwedge\kern-.6pt}%
    {\rule[-\textheight/2]{1ex}{\textheight}}
  }{\textheight}%
}{0.5ex}}%
\stackon[1pt]{#1}{\tmpbox}%
}
\makeatletter

\newcommand{\checknextargt}{\@ifnextchar\bgroup{\gobblenextargt}{}}
\newcommand{\gobblenextargt}[1]{\@ifnextchar\bgroup{, (\ref{#1})\checknextargt}{ and (\ref{#1})\xspace}}
\makeatother
\makeatletter

\newcommand{\checknextargp}{\@ifnextchar\bgroup{\gobblenextargp}{}}
\newcommand{\gobblenextargp}[1]{\@ifnextchar\bgroup{, \ref{#1}\checknextargp}{ and \ref{#1})\xspace}}
\makeatother

\usepackage{array}

\newcommand{\rowstyle}[1]{\gdef\currentrowstyle{#1}%
  #1\ignorespaces
}

\shorttitle{\AbacusSummit{}}
\shortauthors{Maksimova et al.}

\begin{document}

\title{\textsc{AbacusSummit}: A Massive Set of High-Accuracy, High-Resolution $N$-Body Simulations}

\author{Nina A. Maksimova}
\affiliation{Center for Astrophysics $|$ Harvard \& Smithsonian, 60 Garden St., Cambridge, MA 02138}
\author[0000-0002-9853-5673]{Lehman H. Garrison}
\affiliation{Center for Computational Astrophysics, Flatiron Institute, 162 5th Ave., New York, NY 10010}
\author{Daniel J. Eisenstein}
\affiliation{Center for Astrophysics $|$ Harvard \& Smithsonian, 60 Garden St., Cambridge, MA 02138}
\author{Boryana Hadzhiyska}
\affiliation{Center for Astrophysics $|$ Harvard \& Smithsonian, 60 Garden St., Cambridge, MA 02138}
\author{Sownak Bose}
\affiliation{Center for Astrophysics $|$ Harvard \& Smithsonian, 60 Garden St., Cambridge, MA 02138}
\affiliation{Institute for Computational Cosmology, Department of Physics, Durham University, Durham DH1 3LE, UK}
\author{Thomas P.\ Satterthwaite}
\affiliation{Center for Astrophysics $|$ Harvard \& Smithsonian, 60 Garden St., Cambridge, MA 02138}

\correspondingauthor{Lehman Garrison}
\email{lgarrison@flatironinstitute.org}


\begin{abstract}
We present the public data release of the \textsc{AbacusSummit} cosmological $N$-body simulation suite, produced with the \textsc{Abacus} $N$-body code on the Summit supercomputer of the Oak Ridge Leadership Computing Facility.  \textsc{Abacus} achieves $\mathcal{O}\left(10^{-5}\right)$ median fractional force error at superlative speeds, calculating 70M particle updates per second per node at early times, and 45M particle updates per second per node at late times. The simulation suite totals roughly 60 trillion particles, the core of which is a set of 139 simulations with particle mass $2\times10^{9}\,h^{-1}\mathrm{M}_\odot$ in box size $2\,h^{-1}\mathrm{Gpc}$.  The suite spans 97 cosmological models, including Planck 2018, previous flagship simulation cosmologies, and a linear derivative and cosmic emulator grid.  A sub-suite of 1883 boxes of size $500\,h^{-1}\mathrm{Mpc}$ is available for covariance estimation. \textsc{AbacusSummit} data products span 33 epochs from $z=8$ to $0.1$ and include lightcones, full particle snapshots, halo catalogs, and particle subsets sampled consistently across redshift.  \textsc{AbacusSummit} is the largest high-accuracy cosmological $N$-body data set produced to date.

\medskip
\end{abstract}
\keywords{cosmology: theory --- methods: numerical}

\section{\label{sec:level1} Introduction}

In the next decade, large-scale structure surveys such as DESI \citep{Levi:2013gra}, Euclid \citep{2011arXiv1110.3193L}, WFIRST \citep{2015arXiv150303757S}, LSST \citep{2009arXiv0912.0201L}, and the Subaru Hyper Suprime-Cam \citep{2018PASJ...70S...4A} and PFS \citep{Tamura:2016wsg} will catalogue tens of millions of galaxies, mapping an unprecedentedly large volume of space. Extracting sub-percent comparisons between survey observations and cosmological predictions, controlling these surveys' systematic errors, and testing their analysis pipelines for cosmological parameter estimation and covariance requires high-precision mock data.  Although approximate methods are capable of generating sample data, 
cosmological $N$-body simulations are a central tool for generating
accurate forecasts for the non-linear regime of gravitational structure
formation.

Remarkable improvements both in the availability and the performance of computational resources have led to a large increase in the quality and accuracy of cosmological $N$-body simulations. Advances in algorithmic design and high-performance computing have allowed $N$-body codes to simulate more and more particles to better and better mass resolution. 
Roughly speaking, existing $N$-body suites tend to fall in one of two categories: (1) simulations with a large particle count in one cosmology (up to roughly a few trillion particles) e.g.~the Euclid Flagship simulations \citep{2016ascl.soft09016P},  OuterRim \citep{Habib_2016,Heitmann_2019}, Millenium XXL  \citep{2012MNRAS.426.2046A},  DarkSky \citep{skillman2014dark}, Uchuu \citep{2020arXiv200714720I}, TianNu \citep{2017RAA....17...85E},  Horizon \citep{2009ApJ...701.1547K} or (2) larger suites with a smaller number of particles per box, spanning a broader set of cosmologies, such as MultiDark \citep{Klypin_2016}, CoyoteUniverse \citep{Heitmann_2009}, Quijote \citep{Villaescusa_Navarro_2020}, MICE \citep{10.1111/j.1365-2966.2009.16194.x}, AbacusCosmos \citep{GarrisonEtal2017}, DarkQuest \citep{2019ApJ...884...29N}, and Aemulus \citep{2019ApJ...875...69D}.

In anticipation of the forthcoming next generation of large-scale structure surveys, we are releasing a massive suite of high-accuracy $N$-body simulations produced with the \Abacus{} code on the Summit supercomputer at Oak Ridge National Laboratory. \Abacus is a cosmological $N$-body code that is capable of performing trillions of direct pairwise force calculations per second per node with a median $\mathcal{O}\left(10^{-5}\right)$ fractional error on force accuracy. \Abacus{} owes its superlative performance to two key features: (1) a novel algorithm that solves the Poisson equation to great accuracy and (2) performance-optimized, GPU-accelerated software engineering.  \Abacus{} is described in a companion paper to this article (Garrison et al.); previous works on \Abacus include \cite{GarrisonEtal2016}, \cite{GarrisonEtal2017}, and \cite{GarrisonEtal2018}

This simulation suite, named \AbacusSummit{}, consists of 150 simulation boxes, spanning 97 cosmological models, most of which have 330 billion particles with mass $2\times 10^{9}\hMsun$. 
Another 26 boxes provide re-simulation at worse mass resolution for resolution studies and prototyping, and we provide 1883 smaller boxes at the base resolution for investigation of statistical errors.  In total, the suite comprises nearly 60 trillion particles.
\AbacusSummit{} delivers a diversity of publicly available data products designed to cater to a wide variety of science applications, including full particle snapshots, particle subsamples, halo catalogues, particle lightcones, projected density maps, and  particle IDs for forming high-accuracy halo merger trees.

The structure of this paper is as follows: in Section \ref{sec:AbacusSummit}, we lay out the suite specifications of \AbacusSummit, including the relevant code parameters, data products for each simulation, the chosen cosmologies, and the box realizations. In Section \ref{sec:Abacus}, we describe the underlying methods used to produce \AbacusSummit{}: the \Abacus{} $N$-body code, its force solver algorithm, serial and parallel implementations, custom halo-finder CompaSO, on-the-fly lightcone implementation, as well as timestepping, data processing, and force softening methods.
In Section \ref{sec:Performance}, we provide a technical review of the performance of \Abacus on Summit.  In Section \ref{sec:Results}, we present an overview of cosmological opportunities with the simulations.  We summarize in Section \ref{sec:conclusion}.

This article is being published as part of a series of papers describing \Abacus and \AbacusSummit.  The \Abacus code is described in \citet{AbacusCode}, the CompaSO halo finder in \citet{CompaSO}, and the merger trees in \citet{MergerTree}.  With this paper, we are releasing the AbacusSummit simulations for unrestricted public use.


\section{\AbacusSummit: Suite Specifications}\label{sec:AbacusSummit}
\subsection{Overview}

\begin{figure*}[p]
    \centering
    \includegraphics[width=\textwidth]{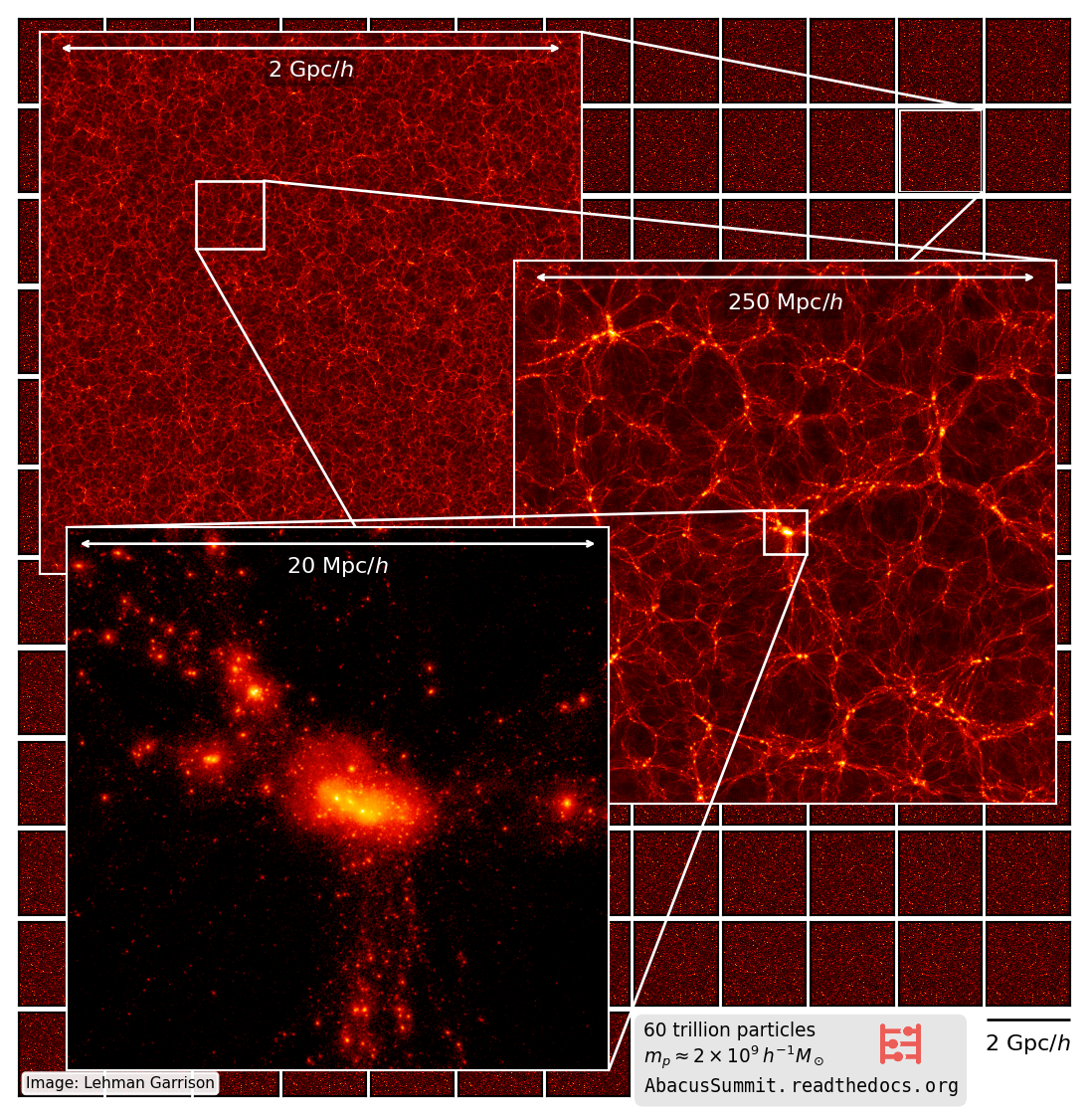}
    \caption{A visualization of the AbacusSummit base-resolution boxes, showing progressive zoom-ins from the full box down to the cluster scale.  The 139 boxes that comprise the base-resolution suite are shown as tiles in the background.  The renderings display the \code{AbacusSummit\_base\_c000\_ph000} simulation at $z=0.1$, and projections are 10 Mpc/$h$ deep.}
    \label{fig:visualization}
\end{figure*}

The \AbacusSummit suite was designed to support a broad range of science applications. This directly informed the choices of cosmological parameters, individual simulation realizations, and data products that comprise the suite. Our main goals were to produce a set of high-accuracy, large-volume simulations of Planck2018 $\Lambda$CDM \citep{Aghanim:2018eyx}, to generate a broad simulation set in secondary cosmologies for exploring cosmological parameter space, and to provide additional simulations sets designed for comparing $N$-body results across different code implementations, resolutions, and/or parameter choices.  A visual summary of a large subset of the simulations is given in Fig.~\ref{fig:visualization}.

\AbacusSummit begins with a large volume (400\,\hGpcC) of high-accuracy simulations of the Planck2018 $\Lambda$CDM cosmological model, separated into 25 base simulations (each of box side $2\,\hGpc$).  These simulations each have $6912^3\approx 330$~billion particles with a particle mass of $2\times 10^9\,\hMsun$ and force softening of $7.2\,h^{-1}$ proper kpc.  We also used this cosmological model for our mass resolution and volume studies.  To study group finding and its dependence on mass resolution, we provide a $6\times$ higher mass resolution simulation in a 1\,\hGpc\ box.  On the other end of the resolution spectrum, we provide two boxes of $7.5\,\hGpc$ size with $27\times$ lower mass resolution simulation in order to output full-sky light cones to $z>2$. 

To explore cosmological-model dependence, we produced simulations of 96 other cosmologies, nearly
all at the base box size and mass resolution.  For four secondary cosmologies, we provide 6 simulations each, phase-matched to the first 6 of the primary cosmology boxes.  We provide 8 simulations matched to the cosmological models of other flagship simulations, as well as 5 simulations chosen to explore our treatment of massive neutrinos.  Finally, we provide simulations of 79 other cosmologies, all matched to the first of the primary cosmology boxes, to support interpolation in an 8-dimensional parameter space, including $\omega_0$, $\omega_a$, $N_{\textrm{eff}}$, and running of the spectral index. 


We provide a suite of 1883 small (500\hMpc) boxes at the base mass resolution to support studies of statistical errors and covariance matrix estimation under periodic boundary conditions. Finally, \AbacusSummit includes 6 simulations with fixed-amplitude white noise in small boxes ($1185\,\hMpc$, with the base mass resolution), with two of those in a fixed-and-paired doublet \citep{Angulo_Pontzen_2016}.

Beyond \AbacusSummit proper, we ran several scale-free and high-redshift simulations.  The scale-free cosmologies span spectral indices $n_s=-1.5$, $-2$, $-2.25$, and $-2.5$, using $N=4096^3$ except for the steepest spectral index, which used $N=6144^3$.  The normalization and output choices for these simulations follow \cite{2021MNRAS.501.5051J}.  The high-redshift simulations all employed $N=6144^3$ in three different box sizes, $20$, $80$ and $300 \hMpc$, terminating at $z=12$, $8$, and $3.5$, respectively, making them well-suited for reionization studies.  Analysis of these simulations will be presented in future work.

For each simulation, we output extensive data products, including particle subsamples, halo catalogs, merger trees, kernel density estimates, light cones, and projected density maps of the light cone. The details of the available data products are described in Section ~\ref{sec:DataProducts}. 

The \AbacusSummit simulation suite was produced using  \Abacus on the Summit supercomputer at the Oak Ridge Leadership Computing Facility (OLCF). Totaling nearly $60$ trillion particles simulated, \AbacusSummit is the largest high-accuracy cosmological N-body data-set produced to date.  With this paper, we are placing the data set into the public domain.

Data access is available through OLCF's Constellation portal.  The persistent DOI describing the data release is \href{https://www.doi.org/10.13139/OLCF/1811689}{10.13139/OLCF/1811689}.  Instructions for accessing the data are given at \url{https://abacussummit.readthedocs.io/en/latest/data-access.html}.

In this Section, we describe the contents of \AbacusSummit, including its full set of cosmologies, individual box realizations, and available data products, as well as notable code choices such as in the generation of initial conditions, timestepping, and force softening. 

\subsection{\AbacusSummit Cosmologies}\label{sec:Cosmologies}
\begin{table*}[tb]
\begin{center}
\footnotesize
\setlength\tabcolsep{3pt}
\begin{tabular}{clccccccccccccc}
Name & Description & $\omega_b$ & $\omega_{c}$ & $h$ & $10^9A_s$ & $n_s$ & $\alpha_s$ & $N_{ur}$ & $N_\mathrm{ncdm}$ &	$10^4\omega_{ncdm}$ &	$w_{0,fld}$ &	$w_{a,fld}$ &	$\sigma_{8,m}$ &	$\sigma_{8,cb}$ \\
    \midrule
\code{cosm000}    & Baseline $\Lambda$CDM & 0.02237 &  0.1200   & 0.6736 & 2.0830   & 0.9649 & 0.0      &   2.0328 & 1      &  6.4420 & -1.0   & 0.0    & 0.807952 & 0.811355                          \\
\code{cosm001}    & Low $\omega_c$ $\Lambda$CDM                                 & 0.02242 &  0.1134   & 0.7030 & 2.0376   & 0.9638 & 0.0      &   2.0328 & 1      &  6.4420 & -1.0   & 0.0    & 0.776779 & 0.780222  \\
\code{cosm002}    & Thawing dark energy   & 0.02237 &  0.1200   & 0.6278 & 2.3140  & 0.9649 & 0.0      &   2.0328 & 1      &  6.4420 & -0.7   & -0.5    & 0.808189 & 0.811577  \\
\code{cosm003}    & $N_{\textrm{eff}}=3.70$        & 0.02260 &  0.1291   & 0.7160 & 2.2438   & 0.9876 & 0.0      &   2.6868 & 1      &  6.4420 & -1.0   & 0.0    & 0.855190 & 0.858583  \\
\code{cosm004}    & $\sigma_{8,m} = 0.75$ $\Lambda$CDM                   & 0.02237 &  0.1200   & 0.6736 & 1.7949   & 0.9649 & 0.0      &   2.0328 & 1      &  6.4420 & -1.0   & 0.0    & 0.749999 & 0.753159  \\
\hline
\end{tabular}
\end{center}
\caption{Cosmological parameters for the first 5 cosmologies; the remaining 94 are available as an electronic table.
}
\label{tab:cosmo}
\end{table*}

\begin{figure*}[p]
\centering
\includegraphics[width=\textwidth]{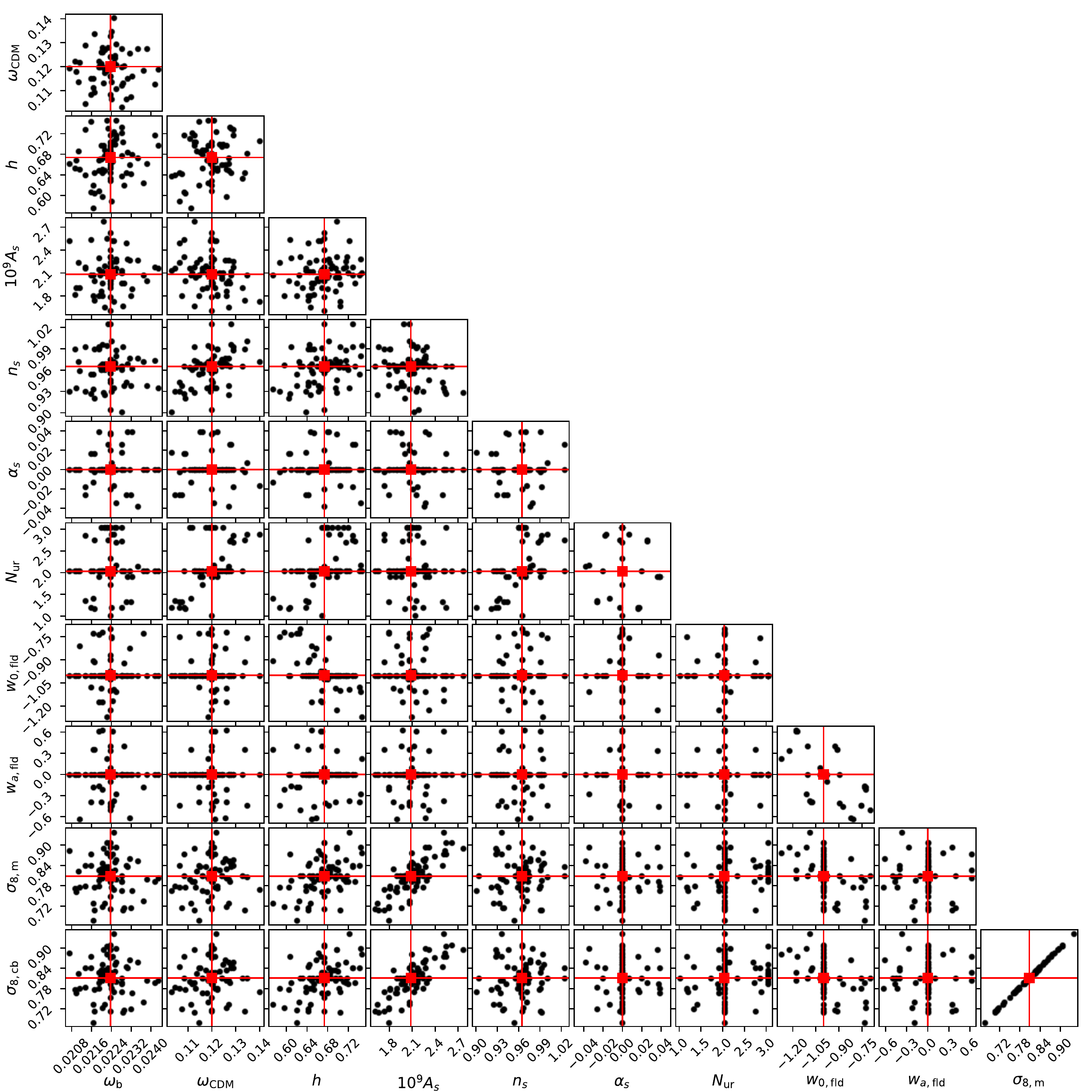}

\caption{A corner plot representation of the cosmologies spanned by \AbacusSummit. The red square marks the fiducial base cosmology (\code{c000}).
}
\label{fig:corner}
\end{figure*}

The cosmologies represented in \AbacusSummit are specified as a set of parameters that are then used to create CLASS input files and \code{abacus.par} parameter files for each simulation run.  
All cosmologies use $\tau=0.0544$ and use \textsc{HyRec} to model recombination. 
The simulations largely span an 8-dimensional cosmological parameter space, parameterized as the cold dark matter density, the baryon density, the Hubble constant, the spectral tilt, the amplitude of structure (i.e., $\sigma_8$), the equation of state of dark energy $w(z)=w_0+(1-a)w_a$, the running of the spectral tilt, and the density of massless relics $N_{\rm eff}$.  All simulations use a flat spatial curvature.  In most cases, $H_0$ is chosen to match the CMB acoustic scale $\theta_*$ ($100 \theta_* = 1.041533$).

Most of the cosmological models include a single species $60$ meV neutrinos, such that $\Omega_{\nu} = 0.00064420$.  A few models use zero or two such species, and one uses a 100 meV species to match an example in the MassiveNuS suite \citep{2018JCAP...03..049L}.
We model massive neutrinos as a smooth component, including their effect in the Hubble expansion rate but ignoring the gravitational forces from clustered neutrinos.
We set the initial conditions by using CLASS to predict the $z=1.0$ power spectrum, combining cold dark matter and baryons, and then scale that power spectrum back to the initial redshift of $z=99$ using the linear growth function including the suppression of growth from the smooth massive neutrino component.
In this way, the simulation will arrive at $z=1$ with the correct linear theory power spectrum for the non-neutrino component.  As we limit our suite to low neutrino masses, the free-streaming scale is large, such that treating this component as smooth is a good approximation on small scales and earlier times.  Effectively our approximation is to slightly overnormalize the large scales in the initial conditions and then have those scales grow mildly slower than the correct dynamics with clustered neutrinos, arriving at $z=1$ with the correct linear-regime power.  Meanwhile, on small scales where the neutrinos are smoother, except at low redshift around the most massive clusters, our approximation will be more correct.


The full range of \AbacusSummit simulated cosmologies is illustrated in Fig.~\ref{fig:corner} and enumerated in Table \ref{tab:cosmo}.  Each cosmology has an identifier with three digits, such as \code{c000} for the Planck 2018 cosmology.  This identifier is included in the simulation name.  Different realizations of the same cosmology are similarly labeled by a three-digit or four-digit phase number, included as \code{ph000} in the simulation name.

The cosmologies are summarized as follows:

\code{c000}: Our chosen fiducial cosmology, matching the Planck 2018 results \citep{Aghanim:2018eyx}, specifically the mean estimates of the TT,TE,EE+lowE+lensing likelihood chains. In this cosmology, we have run twenty-five base mass resolution boxes, the 1883 small boxes, and an additional several boxes at both lower and higher resolutions. \code{c009} is the same cosmology, but with massless neutrinos.

\code{c001-c004}: Four secondary cosmologies, each with six base boxes, including a low $\omega_{c}$ based on WMAP9 + ACT + SPT \citep{2017PhRvD..95f3525C},
a thawing dark energy model ($w_0=-0.7$, $w_a=-0.5$), a model with extra relativistic density ($N_{\textrm{eff}}=3.7$, taken from the \code{base\_nnu\_plikHM\_TT\_lowl\_lowE\_Riess18\_post\_BAO} chain of Planck2018), and a model with lower amplitude clustering (\code{c000} but with $\sigma_8=0.75$).

\code{c010}: Reference cosmology for our previous AbacusCosmos suite \citep{GarrisonEtal2017}.

\code{c012-c018}: Reference cosmologies chosen to match existing flagship simulations and allow for future code comparisons, such as the Euclid Flagship runs, OuterRim, DarkSky, Horizon, IllustrisTNG, and MultiDark \citep{Habib_2016,Heitmann_2019,skillman2014dark,Dubois_2016,nelson2019illustristng,2016MNRAS.457.4340K}.

\code{c019-c020}: Variations around \code{c000} using zero or two massless neutrinos, varying $H_0$ so as to hold the CMB acoustic scale $\theta_*$ fixed.

\code{c021-c022}: Matches to two MassiveNUs simulations \citep{2018JCAP...03..049L}, specifically the massless and single 100~meV models.

\code{c100-c116}: A linear derivative grid with 8 matched pair cosmologies, with small positive and negative steps away from \code{c000} in an 8-dimensional parameter space, as well as one additional model \code{c116} that is the high $\sigma_8$ matched pair to the low $\sigma_8$ secondary cosmology \code{c004}. 
The cosmological model space is parameterized as the cold dark matter density ($\omega_c = \Omega_c h^2$), the baryon density ($\omega_b = \Omega_b h^2$), the spectral tilt $n_s$, the amplitude of structure $\sigma_8$ of the CDM and baryons, the equation of state of dark energy $w(z)=w_0+(1-a)w_a$, the running of the spectral tilt $\alpha_s$, and the density of massless relics $N_{\rm eff}$.  
In detail, we vary $w_a$ holding the equation of state at $z=0.333$ constant, so that a change in $w_a$ is accompanied by a shift in $w_0$ that is --1/4 as big; this was done to allow larger variations in $w_a$ while keeping the low-redshift cosmic distance scale less changed.  Similarly, changes in $N_{\rm eff}$ are compensated by changes in $\omega_c$ and $n_s$, based on the covariances in the Markov Chain posteriors in Planck, so as to leave the CMB less changed.
All simulations use a flat spatial curvature, and $H_0$ is chosen to match the CMB acoustic scale $\theta_*$ to that of \code{c000}.  

\code{c117-c126}: Another linear derivative grid with smaller steps, in $\omega_c$, $n_s$, $w_0$, $w_a$, and $\sigma_8$, to provide more exact linear derivatives for small alterations of the base
\code{c000} model.

\code{c130-c181}: A larger unstructured emulator grid that provides a wider coverage of the 8-dimensional parameter space.  In detail, this was done by placing the points on the surface (not interior) of an 
8-dimensional ellipsoid, whose extent was chosen to be relatively generous (5 to 8 standard deviations in any one parameter) beyond today's constraints from the combination of CMB and large-scale structure data.
After the point on the ellipsoid is set, an additional uniform random excursion between --0.06 and +0.06 is added to $\sigma_8$; this direction was seen as special as other aspects of a cosmological model (such as the optical depth $\tau$) might impact the amplitude of low-redshift clustering while not altering the shape of the power spectrum as much.

The spread of points on the 8-dimensional ellipsoid was constructed to spread the points out, subject
to certain constraints, and to avoid antipodal reflections that would by parity not add new information to the computation of second derivatives.  
Because emulators might not want to use the full 8-dimensional space, we opt to add constraints so that smaller subsets would explore only simpler spaces.  There are four sets of points.
The first varies only $\omega_c$, $\omega_b$, $n_s$, and $\sigma_8$; the second adds $w_0$ and $w_a$; the third adds $\alpha_s$ and $N_{\rm eff}$; and the fourth uses all 8-dimensions.  
In the first set, we do positive and negative excursions purely in the $\omega_c$, $n_s$, and $\sigma_8$ directions and then have 11 additional points.  
We constructed a glass on a unit sphere, subject to these constraints, by starting with random points on the unit sphere, with their antipodes included, and then evolving by an electrostatic repulsion.  The antipodes were discarded in all but three pairs that were held at the unit vector in the $\omega_c$, $n_s$, and $\sigma_8$ directions.  These three pairs were also excluded from the extra uniform spread in $\sigma_8$.

\subsection{\AbacusSummit Simulations}\label{sec:sims}

For \AbacusSummit, we then performed simulations of these cosmological models in a variety of configurations.  For each box size, we may perform simulations with multiple realizations of the initial conditions.  The random seed of the white noise of the Gaussian random field is labeled by \code{ph000-ph4999}.  Hence, simulations with the same phase seed but with different mass resolution or cosmological parameters will share the same initial conditions white noise; such simulations are always of the same box size.  Matching the simulation initial conditions in white noise allows comparison between cosmologies and across mass resolutions with substantially suppressed sample variance.  In particular, this facilitates the construction of derivatives of observables with respect to cosmological parameters.

\code{base}: The bulk of the computational resource is in this configuration, $6912^3$ particles in $2\hGpc$ box.  For \code{c000}, this corresponds to a particle mass of $2\times10^9\hMsun$; this mass will vary slightly in other cosmologies.  There are 139 of these simulations.

\code{high}: A single simulation with 6 times better mass resolution, about $3\times10^8\hMsun$, using $6300^3$ particles in a $1\hGpc$ box.

\code{highbase}: Three simulations with the base particle mass but in $1\hGpc$, hence $3456^3$ particles.  

\code{huge}: Two simulations with $8640^3$ particles in $7.5\hGpc$ boxes, implying 27-fold worse particle mass resolution, around $5\times10^{10}\hMsun$.

\code{hugebase}: Twenty-five simulations of $2\hGpc$ boxes with $2304^3$ particles, matching the mass resolution of \code{huge}.

\code{small}: Simulations of the base mass resolution in $500\hMpc$ boxes, with $1728^3$ particles.  2000 of these simulations were initiated, but some crashed for reasons unrelated to the density field.  We present 1883 of these boxes, 1643 of which reached the final time.

\code{fixedbase}: Simulations of $4096^3$ particles in $1.18\hGpc$ boxes, matching the mass resolution of \code{base}, with white noise chosen to be of fixed amplitude in Fourier space.  That is, all Fourier modes have a random phase but a complex norm of unity, multiplied by the power spectrum.  There are 6 such simulations, 2 of which share identical white noise but with opposite signs for the mode amplitudes.

The individual simulation realizations comprising the \AbacusSummit suite are enumerated in Tables~\ref{tab:simsLCDM} and \ref{tab:sims}. 

\begin{table*}\centering
  \begin{tabular}{>{\raggedright\arraybackslash}l%
   >{\centering\arraybackslash}c%
   >{\centering\arraybackslash}c%
   >{\centering\arraybackslash}c%
   >{\centering\arraybackslash}c%
   >{\raggedright\arraybackslash}p{3cm}%
  }
  \rowstyle{\bfseries}
    Name  & \code{PPD} & Size ($\mathrm{Mpc}/h$)  & $z_{\textrm{final}}$ & Full Outputs & Description \\
    \hline

\code{AbacusSummit\_base\_c000\_ph000}              & 6912 & 2000           & 0.1 & Full                    & Planck2018 $\Lambda$CDM \\
\code{AbacusSummit\_base\_c000\_ph\{001-005\}}      & 6912 & 2000           & 0.1 & Partial+HiZ             & Planck2018 $\Lambda$CDM \\
\code{AbacusSummit\_base\_c000\_ph\{006-024\}}      & 6912 & 2000           & 0.1 & none                    & Planck2018 $\Lambda$CDM \\
\code{AbacusSummit\_high\_c000\_ph100}              & 6300 & 1000           & 0.1 & Full                    & High-res $\Lambda$CDM \\
\code{AbacusSummit\_highbase\_c000\_ph100}          & 3456 & 1000           & 0.1 & Full                    & Base-res $\Lambda$CDM \\
\code{AbacusSummit\_huge\_c000\_ph\{201-202\}}      & 8640 & 7500           & 0.1 & 1.4, 1.1, 0.8, 0.5, 0.2 & Low-res $\Lambda$CDM, box-centered lightcone \\
\code{AbacusSummit\_hugebase\_c000\_ph\{000-024\}}  & 2304 & 2000           & 0.1 & 1.4, 1.1, 0.8, 0.5, 0.2 & Low-res match to base \\
\code{AbacusSummit\_fixed\_c000\_ph099}          & 4096 & 1185           & 0.1 & Full                    & Base-res $\Lambda$CDM, fixed amplitude \\
\code{AbacusSummit\_fixed\_c000\_ph098}          & 4096 & 1185           & 0.1 & Full                    & Base-res $\Lambda$CDM, phase inverted from ph099 \\
\code{AbacusSummit\_small\_c000\_ph\{3000-4999\}}      & 1728 & 500           & 0.2 & none                    & Base-res $\Lambda$CDM small boxes \\
\hline
  \end{tabular}
  \caption{The \AbacusSummit simulations in the base Planck2018 $\Lambda$CDM cosmology (\code{c000}).  The simulation name contains the label corresponding to the specific phase realization of each cosmology (e.g. \code{ph007}). \code{PPD} refers to the particles per dimension; \code{PPD}$^3$ gives the total number of particles simulated in the box. The length of the cubic box is specified in the ``Size'' column. Only a few of our simulations include the full timeslice; typically only subsamples are saved. The ``Full Outputs'' column states the redshifts at which timeslices were saved (in addition to subsamples). ``Full'' refers to the full list of output redshifts, defined as z=3.0, 2.5, 2.0, 1.7, 1.4, 1.1, 0.8, 0.5, 0.4, 0.3, 0.2, 0.1. The ``Partial'' list is z = 2.5, 1.4, 0.8, 0.2. ``Partial+HiZ'' adds z=3.0 and 2.0 to that.}
  \label{tab:simsLCDM}
\end{table*}

\begin{table*}\centering
  \begin{tabular}{>{\raggedright\arraybackslash}l%
   >{\centering\arraybackslash}c%
   >{\centering\arraybackslash}c%
   >{\centering\arraybackslash}c%
   >{\centering\arraybackslash}c%
   >{\raggedright\arraybackslash}p{3cm}%
  }
  \rowstyle{\bfseries}
    Name  & \code{PPD} & Size ($\mathrm{Mpc}/h$)  & $z_{\textrm{final}}$ & Full Outputs & Description \\
    \hline

\code{AbacusSummit\_base\_c001\_ph000}              & 6912 & 2000           & 0.1 & Partial+HiZ             & WMAP7 model with low $\Omega_c$ \\
\code{AbacusSummit\_base\_c001\_ph\{001-005\}}      & 6912 & 2000           & 0.1 & Partial  & WMAP7 \\
\code{AbacusSummit\_fixed\_c001\_ph099}          & 4096 & 1185           & 0.1 & Full                    & WMAP7, fixed amplitude \\
\hline

\code{AbacusSummit\_base\_c002\_ph000}              & 6912 & 2000           & 0.1 & Partial+HiZ             & thawing wCDM $w_0 = -0.7$, $w_a = -0.5$ \\
\code{AbacusSummit\_base\_c002\_ph\{001-005\}}      & 6912 & 2000           & 0.1 & Partial                 & thawing wCDM \\
\code{AbacusSummit\_fixed\_c002\_ph099}          & 4096 & 1185           & 0.1 & Full                    & thawing wCDM, fixed amplitude \\

\hline

\code{AbacusSummit\_base\_c003\_ph000}              & 6912 & 2000           & 0.1 & Partial+HiZ             & $N_{\rm eff}=3.70$ model \\
\code{AbacusSummit\_base\_c003\_ph\{001-005\}}      & 6912 & 2000           & 0.1 & Partial                 & $N_{\rm eff}=3.70$ model \\
\code{AbacusSummit\_fixed\_c003\_ph099}          & 4096 & 1185           & 0.1 & Full                    & $N_{\rm eff}=3.70$ model, fixed amplitude \\

\hline

\code{AbacusSummit\_base\_c004\_ph000}              & 6912 & 2000           & 0.1 & Partial+HiZ             & Low $\sigma_{8,\textrm{matter}}$ = 0.75, otherwise baseline $\Lambda$CDM \\
\code{AbacusSummit\_base\_c004\_ph\{001-005\}}      & 6912 & 2000           & 0.1 & Partial                 & Low $\sigma_{8,\textrm{matter}}$ = 0.75 $\Lambda$CDM \\
\code{AbacusSummit\_fixed\_c004\_ph099}          & 4096 & 1185           & 0.1 & Full                    & $\sigma_{8,\textrm{matter}}$ = 0.75 $\Lambda$CDM, fixed amplitude \\

\hline

\code{AbacusSummit\_base\_c009\_ph000}              & 6912 & 2000           & 0.1 & Partial                 & Baseline $\Lambda$CDM with massless neutrinos \\
\code{AbacusSummit\_base\_c010\_ph000}              & 6912 & 2000           & 0.1 & none                    & AbacusCosmos LCDM \\
\code{AbacusSummit\_base\_c012\_ph000}              & 6912 & 2000           & 0.1 & none                    & Euclid Flagship1 \\
\code{AbacusSummit\_base\_c013\_ph000}              & 6912 & 2000           & 0.1 & none                    & Euclid Flagship2 \\
\code{AbacusSummit\_base\_c014\_ph000}              & 6912 & 2000           & 0.1 & none                    & OuterRim \\
\code{AbacusSummit\_base\_c015\_ph000}              & 6912 & 2000           & 0.1 & none                    & Dark Sky \\
\code{AbacusSummit\_base\_c016\_ph000}              & 6912 & 2000           & 0.1 & none                    & Horizon \\
\code{AbacusSummit\_base\_c017\_ph000}              & 6912 & 2000           & 0.1 & none                    & Illustris \\
\code{AbacusSummit\_base\_c018\_ph000}              & 6912 & 2000           & 0.1 & none                    & Multidark Planck \\
\hline

\code{AbacusSummit\_base\_c019\_ph000}              & 6912 & 2000           & 0.1 & none                    & Baseline $\Lambda$CDM w/two 60 meV neutrino species \\
\code{AbacusSummit\_base\_c020\_ph000}              & 6912 & 2000           & 0.1 & none                    & Baseline $\Lambda$CDM w/massless neutrinos \\
\code{AbacusSummit\_highbase\_c021\_ph000}              & 3456 & 1000           & 0.1 & Partial                    & MassiveNUs base model with massless neutrinos \\
\code{AbacusSummit\_highbase\_c021\_ph000}              & 3456 & 1000           & 0.1 & Partial                    & MassiveNUs with 100 meV neutrino species \\
\hline

\code{AbacusSummit\_base\_c\{100-115\}\_ph000}      & 6912 & 2000           & 0.1 & none                    & Linear Derivative Grid \\
\code{AbacusSummit\_base\_c116\_ph000}              & 6912 & 2000           & 0.1 & none                    & Linear Derivative Grid \\
\code{AbacusSummit\_base\_c\{117-126\}\_ph000}      & 6912 & 2000           & 0.1 & none                    & Finer Linear Derivative Grid \\

\code{AbacusSummit\_base\_c\{130-181\}\_ph000}      & 6912 & 2000           & 0.1 & none                    & Broader Emulator Grid  \\
\hline

  \end{tabular}
  \caption{Like Table~\ref{tab:simsLCDM}, but for simulations with cosmologies other than the baseline \code{c000} choice. The simulation name encodes the cosmology choice (e.g. \code{c001}): a brief description of the cosmology is given in the rightmost column and for more details, refer to Table~\ref{tab:cosmo}.}
  \label{tab:sims}
\end{table*}

\subsection{Initial Conditions}\label{sec:IC}
We generate initial conditions with the public \code{zeldovich-PLT}\footnote{\url{https://github.com/abacusorg/zeldovich-PLT}} code, which uses the first-order particle linear theory corrections described in \cite{GarrisonEtal2016}.  The second-order Lagrangian perturbation theory correction is computed with direct force evaluation during the first two steps of \Abacus.  To aid methods that seek to reduce large-scale structure sample variance by utilizing cross-correlation with the initial conditions, we make available reduced-resolution versions of the initial displacements and linear density field.  These are provided at $576^3$ and $1152^3$ resolutions and only include the first-order (ZA) part.  The lower 10 bits of the displacements and densities are truncated to aid compression, a modification of $< 0.01\%$.  The displacements and densities are stored in separate  files, the headers of which include the simulation parameters, the full CLASS power spectrum, and table of growth factors as computed by \Abacus's cosmology module.

We do not provide the full-resolution initial conditions, but we do make available the \code{abacus.par} parameter file as well as the CLASS input power spectrum \citep{lesgourgues2011cosmic}, from which the initial conditions may be readily regenerated using \code{zeldovich-PLT}. 

As described in \cite{GarrisonEtal2016}, \code{zeldovich-PLT} performs a rescaling to correct for the violation of linear theory that occurs on small scales in discrete particle systems. The target correction redshift is chosen to be $z=12$, and the particles are displaced from an initial cubic grid using the particle lattice eigenmodes. The particles' initial positions relative to the starting grid are permanently encoded as part of their particle IDs.

With the approximation of neutrinos as a smooth, non-clustering component (Sec.~\ref{sec:Cosmologies}), the growth rate, and therefore the linear theory velocities, is modified.  Defining $f_c = 1-\Omega_\nu/\Omega_M$ as the clustering fraction, the growth rate correction can ordinarily be applied in configuration space as a multiplicative prefactor of $\left(\sqrt{1 + 24f_c} - 1\right)/4$ on the velocity.  However, with the PLT correction of a $\mathbf{k}$-dependent eigenvalue, $\epsilon(\mathbf{k})$, the $f_c$ and PLT factors must be combined in Fourier space at the mode level, yielding a prefactor of $\left(\sqrt{1 + 24\epsilon(\mathbf{k})f_c} - 1\right)/4$ on each velocity mode.

\subsection{Code Parameters}\label{sec:Params}

\begin{figure}[t]
\centering
\includegraphics[width=\columnwidth]{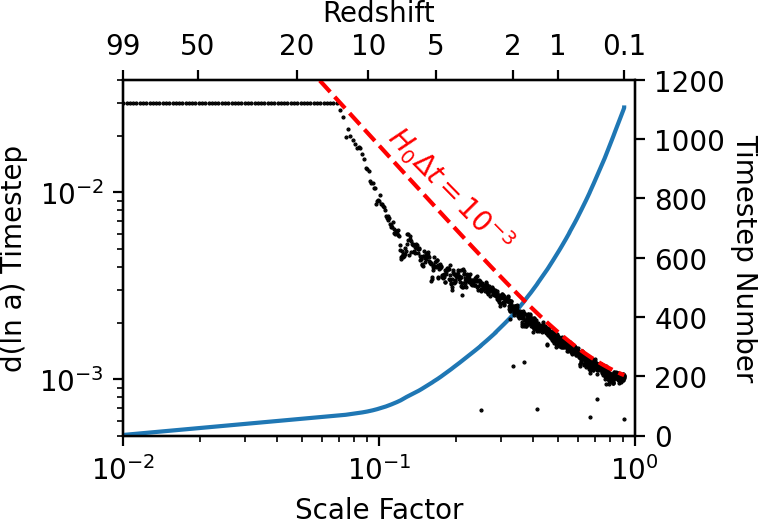}
\caption{The timestep in the \code{base\_c000\_ph000} sim as a function of the scale factor $a$.  Black points show the change in $\ln(a)$, showing that the simulation begins with steps of $\Delta\ln(a)=0.03$ in the pre-collapse regime, and then drops rapidly at $z=15$, finishing with steps around $10^{-3}$.  The few particularly short time steps are due to choosing to arrive at a particular output redshift.  The red line shows a line of constant change in proper time; because we use a proper softening length, the simulation tracks this behavior, with steps of about $0.001H_0^{-1} \approx 10^7h^{-1}$~years.  The blue line and the scale on the right-hand side shows the cumulative number of timesteps as a function of scale factor.
}
\label{fig:timestep}
\end{figure}

\AbacusSummit{} simulations use the spline force softening scheme described in \cite{GarrisonEtal2018}, chosen because the force law returns to exactly the $1/r^2$ form at a scale mildly larger than the softening length.  This is desirable because the far-field method computes unsoftened forces, and because it limits the modification of the force law to scales where the two-body scattering timescale is short.  In \cite{Garrison+2021}, we show that using force softening that is fixed in proper, rather than comoving, distance gives more efficient solutions, reaching mildly better convergence to the small-scale correlation function in scale-free simulation tests while requiring substantially fewer timesteps.  The physical reasoning is clear: the collapsed interiors of halos aren't participating in the Hubble expansion and so one should maintain a consistent force law, consistent with the ideas of stable clustering.  For our base boxes, we select a Plummer-equivalent softening length of $7.2h^{-1}$~proper kpc, capped at 30\% of the interparticle spacing at early times ($z>11$).  Simulations with other mass resolutions scale the softening by the interparticle spacing.  \cite{Garrison+2021} shows that one should expect few per cent convergence on the two-point correlation function at about 10\% of the interparticle spacing, so about $30h^{-1}$~proper kpc. It also demonstrates that this convergence is not improved by yet smaller softening lengths but is instead determined by the mass resolution; we therefore have chosen to avoid the extra computational cost that the smaller timesteps required by finer softening would incur.  Clustering well inside the proper softening length is suppressed relative to if the softening were fixed in comoving distance to the $z=0$ value, but these are precisely the scales for which neither solution is converged due to the mass resolution limit.

As described in further detail in Sec.~\ref{sec:Abacus}, \Abacus subdivides the simulation space into a cubic grid with \code{CPD} cells per dimension. \code{CPD} is chosen solely to balance the computational workload in the near-field and far-field force computations and does not affect the accuracy or physical results of the simulation.  The cell structure does appear in some aspects of the data product file organization, as will be described later.
The \AbacusSummit base simulations set \code{CPD} to 1701, yielding an average of 67 particles per cell. 

\Abacus evolves the particle dynamics using second-order leapfrog integration and  global timestepping, where all particles use the same timestep.  While this is imposed by the current structure of the code, it also offers some important advantages, most notably that since the timestep is chosen to serve the densest regions, most of the particles have a rather short time step and more accurate integrations.  The variation of individualized timesteps also tends to break the symplectic property of leapfrog \citep{1997astro.ph.10043Q}, although more complicated formulations can solve this \citep{Farr2007}; we have not explored whether our global timestepping does better in this regard.  We note that cosmological simulations such as \AbacusSummit{} do not have the mass resolution to properly resolve dynamical resonances, regardless of time stepping \citep[e.g.,][]{2007MNRAS.375..425W}. 

The timestep selection criteria are described in detail in \cite{GarrisonEtal2018}.  For most of the steps, we are limited by our small-scale criteria, which is computed as the minimum value over all cells of the ratio of the per-cell velocity dispersion to the per-cell maximum particle acceleration, scaled by a tuning parameter $\eta$. For \AbacusSummit{}, we choose $\eta$ to be 0.25.  At early times when the particles are all close to the initial grid, we are limited by a cosmological bound that the global time step $d(\ln a)$, for scale factor $a$, never exceed 0.03.

The resulting timestep behavior of a typical base-resolution \AbacusSummit{} simulation (particle mass $2\times 10^9 \hMsun$) is shown in Figure \ref{fig:timestep}.  We require roughly 1100 steps to reach $z=0.1$. The timestep in $\ln a$ decreases rapidly at $z=15$ with the first close approach of pairs of particles (but recall that the simulation is highly softened at early times because of the use of proper softening).  At late times, the timestep asymptotes to a value of $0.001H_0^{-1}\approx 10h^{-1}$~Myr.  This constant level is expected because of the constant proper softening and the fact that the innermost density of halos increases only slowly.  It is perhaps surprising that this constant-time asymptote is approached from below, not above.  We expect this is because the velocity dispersion of the Mpc-scale cell around the dense cores is increasing over time as the bigger halos build up.  The effect is particularly noticeable at redshift 5 to 10, where the collapsed halos are often smaller than a cell, so that the velocity dispersion is diluted by the colder circum-halo particles.


\subsection{Data Products}\label{sec:DataProducts}

\AbacusSummit{} consists of a total of roughly 2 PB of available data products. For the purposes of describing output data products, we define 12 primary redshifts and 21 secondary redshifts between $z = 0.1$ and $z = 8.0$. The primary set is $z=$0.1, 0.2, 0.3, 0.4, 0.5, 0.8, 1.1, 1.4, 1.7, 2.0, 2.5, and 3.0; the secondary set is 0.15, 0.25, 0.35, 0.45, 0.575, 0.65, 0.725, 0.875, 0.95, 1.025, 1.175, 1.25, 1.325, 1.475, 1.55, 1.625, 1.85, 2.25, 2.75, 3.0, 5.0, and 8.0.  We have designed a set of data products aimed at supporting mock catalogs to be constructed using halo occupation distributions, as well as efficient access to measurements of the density fields.

For \AbacusSummit halo catalogues and related data products, \Abacus uses CompaSO: a custom, on-the-fly halo finding algorithm summarized in Sec.~\ref{sec:halofinding} and described comprehensively in \cite{CompaSO}. 
Using CompaSO, we identify and output halo information at 33 epochs.
For the purpose of this section, we note that CompaSO processing results in particles being assigned to disjoint L0 sets, inside of which are found disjoint L1 halos, inside of which are found disjoint L2 subhalos.  The L1 halos are the key CompaSO product and for these we output a large set of summary statistics, listed in Table~\ref{tab:halocat}, as well as constituent particles.

\begin{table*}
  \begin{tabular}{p{6cm} p{11cm}}
\code{uint64\_t id } &  A unique halo number. \\
\code{uint64\_t npstartA } &  Where to start counting in the particle output for subsample A. \\
\code{uint64\_t npstartB } &  Where to start counting in the particle output for subsample B. \\
\code{uint32\_t npoutA } &  Number of taggable particles written out in subsample A. \\
\code{uint32\_t npoutB } &  Number of taggable particles  written out in subsample B. \\
\code{uint32\_t ntaggedA } &  Number of tagged particle PIDs written out in subsample A. \\
\code{uint32\_t ntaggedB} &  Number of tagged particle PIDs written out in subsample B.\\
\code{uint32\_t N } &  The number of particles in this halo. \\
\code{uint32\_t L2\_N[N\_LARGEST\_SUBHALOS] } &  The number of particles in the largest L2 subhalos. \\
\code{uint32\_t L0\_N } &  The number of particles in the L0 parent group. \\
\code{float SO\_central\_particle[3] } &  Coordinates of the SO central particle. \\
\code{float SO\_central\_density } &  Density of the SO central particle. \\
\code{float SO\_radius } &  Radius of SO halo (distance to particle furthest from central particle). \\
\code{float SO\_L2max\_central\_particle[3] } &  Coordinates of the SO central particle for the largest L2 subhalo. \\
\code{float SO\_L2max\_central\_density } &  Density of the SO central particle of the largest L2 subhalo. \\
\code{float SO\_L2max\_radius } &  Radius of SO halo (distance to particle furthest from central particle) for the largest L2 subhalo. \\ \hline
\multicolumn{2}{p {17cm}} {The quantities below are computed using a center defined by the center of mass position and velocity of the largest L2 subhalo. In addition, the same quantities with \code{\_com} use a center defined by the center of mass position and velocity of the full L1 halo. All second moments and mean speeds are computed only using particles in the inner 90\% of the mass relative to this center.} \\\hline
\code{float x\_L2com[3] } &  Center of mass position of the largest L2 subhalo. \\
\code{float v\_L2com[3] } &  Center of mass velocity of the largest L2 subhalo. \\
\code{float sigmav3d\_L2com } &  The 3-D velocity dispersion, i.e., the square root of the sum of eigenvalues of the second moment tensor of the velocities relative to the center of mass. \\
\code{float meanSpeed\_L2com } &  Mean speed of particles, relative to the center of mass. \\
\code{float sigmav3d\_r50\_L2com } &  Velocity dispersion (3-D) of the inner 50\% of particles. \\
\code{float meanSpeed\_r50\_L2com } &  Mean speed of the inner 50\% of particles. \\
\code{float r100\_L2com } &  Radius of 100\% of mass, relative to L2 center. \\
\code{float vcirc\_max\_L2com } &  Max circular velocity, relative to the center of mass position and velocity, based on the particles in this L1 halo . \\
\code{int16\_t sigmavMin\_to\_sigmav3d\_L2com } &  Min(\code{sigmav\_eigenvalue}) / \code{sigmav3d}, condensed to [0,30000]. \\
\code{int16\_t sigmavMax\_to\_sigmav3d\_L2com } &  Max(\code{sigmav\_eigenvalue}) / \code{sigmav3d}, condensed to [0,30000]. \\
\code{uint16\_t sigmav\_eigenvecs\_L2com } &  Eigenvectors of the velocity dispersion tensor, condensed into 16 bits. \\
\code{int16\_t sigmavrad\_to\_sigmav3d\_L2com } &  \code{sigmav\_rad} / \code{sigmav3d}, condensed to [0,30000]. \\
\code{int16\_t sigmavtan\_to\_sigmav3d\_L2com } &  \code{sigmav\_tan} / \code{sigmav3d}, condensed to [0,30000]. \\
\code{int16\_t r\{10,25,33,50,67,75,90, 95,98\}\_L2com} &  Radii of this percentage of mass, relative to L2 center. Expressed as ratios of \code{r100} and condensed to [0,30000]. \\
\code{int16\_t sigmar\_L2com[3] } &  The square root of eigenvalues of the moment of inertia tensor, as ratios to \code{r100}, condensed to [0,30000]. \\
\code{int16\_t sigman\_L2com[3] } &  The square root of eigenvalues of the weighted moment of inertia tensor, in which we have computed the mean square of the normal vector between the COM and each particle, condensed to [0,30000]. \\
\code{uint16\_t sigmar\_eigenvecs\_L2com } &  The eigenvectors of the inertia tensor, condensed into 16 bits. \\
\code{uint16\_t sigman\_eigenvecs\_L2com } &  The eigenvectors of the weighted inertia tensor, condensed into 16 bits. \\
\code{int16\_t rvcirc\_max\_L2com } &  radius of max circular velocity, relative to the L2 center, stored as the ratio to \code{r100} condensed to [0,30000].
  \end{tabular}
  \caption{Data products available for an output redshift of \AbacusSummit, as part of the \textsc{CompaSO} halo catalogue. }
  \label{tab:halocat}
\end{table*}

Beyond halos, we output a 10\% subsample of particles.  This is suitable for many applications, e.g., the computation of matter-field statistics such as gravitational lensing observables and the Monte Carlo sampling of the halo mass distribution for the siting of satellite galaxies in halo-occupation modeling.  The subsample of particles is consistent between output redshifts and the particle ID numbers are included, so that one can track individual particles in time.  At primary redshifts, we output the position, velocity, particle ID, and kernel density estimate of the full subsample of particles.

By connecting the particle ID numbers for a small, consistent subsample of particles in each halo, one can build merger trees by post-processing to find which halos contain matching sets of particles. We provide merger tree aggregations, described in \cite{MergerTree}, that use the IDs to rapidly associate halos between time slices and build up the progenitor history of the late-time halos.

The key data products available as part of \AbacusSummit{} are:

\begin{enumerate}
    \item CompaSO-identified L1 halo catalogues with comprehensive statistics computed on-the-fly from the full particle set, available at all primary and secondary redshifts.
    \item Merger tree associations of these halos, generated from post-processing of the particle IDs between adjacent redshifts. We additionally store halo progenitor information matched across epochs separated by two snapshots, which provides a simple diagnostic for the frequency of transient fly-by events.
    \item Cleaned halo catalogs, row-matched to the L1 catalogues.  The cleaning uses the merger trees to aggregate halos that are presently separate but that were unified in the past; this appears to redress some examples of overly aggressive deblending of halos. 
    \item 10\% subsamples of particle data, split into 3\% and 7\% sets, called A and B, so that users can minimize their data access based on application.  These sets of particles are consistently defined across redshift and are selected effectively randomly based on a hash of the particle ID number.  At primary redshifts, we output the position, velocity, particle ID, and kernel density estimate of the full subsample of particles, split into two sets of files based on whether each particle belongs to an L0 halo or not. For each L0 halo, particles belonging to the same L1 halo are in contiguous sets, and the indices required to access them are stored in the halo catalogue statistics. Particle positions and velocities are output in one file. A separate file contains the unique particle ID, which encodes the initial grid position, as well as the kernel density estimate and a sticky bit that is set if the particle has ever been in the most massive L2 halo of an L1 halo with more than 35 particles.  At secondary redshifts, we provide only the particles in L1 halos and only particle ID file.
    \item  A light cone stretching from the corner of the box and including a single second periodic copy of the box. This provides an octant of sky to $z=0.8$ and about 900 square degrees to $z=2.45$. For particles belonging to both the $10\%$ subsample and the lightcone at any given epoch, we output their positions, velocities, and IDs, as well as each particle's HEALPix pixel number (in Nested order) used to form projected density maps, where $N_{\textrm{side}}=16384$. 
    \item In addition to all the above, we also output the full particle set for a few of the primary timeslices of a few boxes, including all positions, velocities, and particle IDs.
    \item Matter power spectrum measurements using the 10\% particle subsample, or the full particle set when available.  The $\ell=0$ real-space and $\ell=0,2,4$ redshift-space spectra are reported.
    \item The initial density field and corresponding particle displacements at $576^3$ and $1152^3$ resolutions.
\end{enumerate}

The purpose of the secondary redshifts is to support generation of merger trees, as the particle IDs can be used to associate halos between snapshots. Therefore, for the 24 secondary redshifts, we output the halo catalogs and the halo subsample particle IDs (which encode each particle's kernel density estimate and its sticky L2 tag) only, but not the positions or velocities of the halo particles, nor any information regarding the field particles.  

A particle is tagged if it is taggable and is in the largest L2 halo for a given L1 halo. The particle IDs can likewise be used to associate halo catalogues to the light cones, as the particle subsamples are consistent across epochs. We stress likewise that the particle IDs---which are output for the particle subsample at all secondary redshifts and for the full particle set at the primary redshifts---each contain the kernel density estimate, which in its own right forms a powerful data product since it is output for many particles at a large number of redshifts. It is also  possible to use the secondary redshifts to associate halo catalogs to the light cones, as the subsample of particles (and therefore their IDs) are the same.

\subsection{File Formats and Directory Structures}\label{sec:file_formats}

We have organized the data products of \AbacusSummit\ into structures that we believe will be highly useful for users while trying to minimize the data volume.  Even with these aggressive compressions, the total data volume is nearly 2 PB, and so it is important that most user applications not need to read the entire dataset.  More detailed documentation is available at \url{https://abacussummit.readthedocs.io}; here we focus on some design aspects that might be useful for other programs.

Each simulation is located in its own directory, and the bulk of the data is found in four subdirectories: \code{halo} for the halo catalogs and particle subsamples, \code{lightcone} for the lightcone outputs, \code{cleaned} for the post-processed cleaned catalogs, and \code{slices} for the time slices.  The next subdirectory level separates the various redshifts.  Below that, each type of file is given a separate subdirectory, so that a user fetching only particular file types should find them contiguous on a tape archive.  And within each of these subdirectories, the files are split into chunks called \textit{superslabs}, as they are the concatenation of several simulation slabs.  The intention of the superslab division is to allow the user to segment their loading of the simulation for a rolling processing.  For the base simulations, there are 34 superslabs, each comprised of 50 slabs (with the last being 51), so each superslab is a 3\% planar region of the survey volume.  We also provide the parameter file for the simulation invocation, the linear power spectrum file, and a summary file that contains a table of cosmological epoch, timing, and statistics for each time step.

We use ASDF \citep[Advanced Scientific Data Format,][]{GREENFIELD2015240} as our container file format, as this offers human-readable headers while allowing us to separate each column of our catalogs into a separate block.  This allows users to efficiently load subsets of the columns, and we chose an ordering of columns that we hope will yield some contiguous loading patterns.  Further, we chose the size of the superslabs so that the data volumes in individual columns in these files would typically be at least 10 MB, so as to amortize disk seek-time latency.

For the particle subsamples, as many applications only need particles in the halos, we separate the particles into two sets of files for halo and field, based on the L0 segmentation.  It is important to note that while the chunk boundaries of the field file are set simply by the cell boundary, i.e., clean planar cuts, the boundaries of the halo file are set based on whether the halo (not the particle!) belongs to a cell.  Halos frequently span multiple cells, so the halo files do not have a clean planar boundary.  To get a complete union of particles in a small region, one may need to draw from two consecutive halo-set files.  We further separate the halo and field files into a 3\% and 7\% subsample, called A and B, so that users can efficiently access 3\%, 7\%, or 10\% of the particles for their application.

We were aggressive in trimming the bit-level precision of output quantities, as the low-significance bits are of little to no science value but foil data compression algorithms.  Notably for the particle subsamples, we limited positions to 20 bits across the box (about 2 kpc/h granularity and 0.6 kpc/h rms) and velocities to 12 bits from $-6000$ to $+6000$ km/s (3 km/s granularity and 0.9 km/s rms); these were combined into a single 32-bit integer for simple file loading, and therefore we refer to the format as \code{rvint}.  For the full particle time slices, where there are many particles in a single cell, we have a more complicated format called \code{pack9} in which positions and velocities are stored with 12 bits each, but relative to a scaling in a per-cell header.  This gives mildly higher resolution (0.3 kpc/h granularity) while holding the compressed size to 8.5 bytes per particle.

Particle ID numbers and the kernel density estimates are placed in separate order-matched files, using a 64-bit integer.  We store the square-root of the density estimate in units of the cosmic mean as an integer.  We note that the small radius of the kernel is such that even a single particle gives a density of about 10 in these units; i.e., the density is useful in halo work, but is not a good measure of densities near the cosmic mean. 

Halo catalog columns were carefully considered for suppression of bit precision.  For example, radii are reported as a ratio to the maximum particle radius in the halo, because this ratio falls in the range 0 to 1, we compressed it to a 16-bit unsigned integer.  Similar ratios are used for velocity dispersions relative to the 3-dimensional dispersion, which provides a robust maximum.  The rank-3 moment of inertia tensors were diagonalized so that the eigenvalues could be outputted as ratios and the three Euler angles of the eigenvectors compressed heavily into a single 16-bit integer giving few degree precision. We provide a Python reader, distributed on GitHub and PyPI as \texttt{abacusutils}\footnote{\url{https://github.com/abacusorg/abacusutils}}, that can read subsets of columns and that performs a translation of all of these ratios and other compressions back into physical units.

After this explicit truncation of precision, we then perform lossless data compression with the \code{zstd}\footnote{\url{https://facebook.github.io/zstd/}} algorithm in the \code{Blosc} package\footnote{\url{https://www.blosc.org/}}.
Prior to compression, we perform transposition of the bits or bytes in the data vector, a feature built into \code{Blosc}.  In other words, if one has 1 million 4-byte numbers, byte transposition would yield a vector of the million first bytes, followed by the million second bytes, etc.  Bit transposition would have the million first bits, then the million second bits, etc.  This helps the compression ratios substantially, and we recommend investigating this in any data compression of a vector of fixed-sized numbers.

We were disappointed at how poorly standard algorithms perform at identifying patterns that occur every 4th byte, as commonly happens with integer data that only rarely produce values above $2^{16}$ (but can and therefore can't be shorted to 16 bits) or with floating-point data whose exponents don't explore a wide range.  After bit or byte transpose, these patterns are made contiguous in the file and the compression algorithms squeeze the result to almost nothing.  For example, for our storage of 4-byte healpix pixel numbers (which reach 3 billion), we sorted the values in internally convenient segments of the output and then did this compression; the result requires only 1/3 of the byte per particle.

Rather than transposing an entire column, higher performance can be gained by performing the transposition in cache-sized blocks.  The default \code{Blosc} recommendation is to target the CPU's L1 cache size (typically 32 KB), and while this yields high performance, we found that this small size missed substantial compression opportunities.  We studied the performance/compression tradeoff of various block sizes, and found that a few MB yielded nearly maximal compression while still providing $>500$ MB/s of decompression per core, including both the inverse transpose and the \code{zstd} decompression.  Since the decompression parallelizes over blocks, this was deemed fast enough to keep up with almost any storage medium with only a few cores.

Similarly, we experimented with several backing compressions before settling on \code{zstd}.  \code{lz4} offered a compelling alternative, and indeed was faster than \code{zstd}, but missed substantial compression opportunities that \code{zstd} found even at its lowest compression level.  The resulting smaller file sizes also result in higher performance in the limit of slow storage media.

We provide an extension to the ASDF Python reader that incorporates the \code{Blosc} engine, so that all of the decompression (including the inverse transpose) is invisible to the user.  The extension is transparently installed via the ASDF compression extension mechanism, which was developed in support of this project, when the user installs the \code{abacusutils} Python package.

For base simulations without full time-slice outputs, the total set of data products is about 7.85 TB.  This is about 2.2 TB for the 33 redshift catalogs, 1.3 TB for the lightcone information, and 4.4 TB for the particle subsamples.  When time-slice outputs do exist, they are 3.5 TB per redshift, which is about 10.6 bytes per particle including the particle ID.

All \Abacus data products undergo 32-bit checksum verification to ensure accuracy and guard against corruption as they flow through the \AbacusSummit production pipeline, from production to compression all the way through to data transfer to their final location. We have implemented \code{fast-cksum}\footnote{\url{https://github.com/abacusorg/fast-cksum/}}: a checksum utility that produces identical output to the GNU \code{cksum} utility provided in most Linux environments, but with roughly a factor of 10 improvement in speed.  The performance is obtained from precomputing lookup tables for all possible 16-byte values convolved with the CRC generating polynomial\footnote{This approach was derived from \url{https://github.com/stbrumme/crc32}}.

Using \code{fast-cksum}, \Abacus computes and stores the checksums of all data products on-the-fly before the data is written out to disk. Every subsequent time the data products are read off of disk during post-processing, the \AbacusSummit production pipeline re-computes each file's checksum and verifies it against the recorded original checksum before proceeding to compression. After data compression, we store a new list of checksums for all compressed data product files and verify these, again, during data transfer to NERSC and OLCF HPSS. 



\section{Abacus}\label{sec:Abacus}

\subsection{Force Solver Algorithm and Serial Method}\label{sec:method}

The \AbacusSummit{} suite was executed on the Summit supercomputer using the high-performance, high-accuracy N-body code \Abacus{}. \Abacus{} uses a variant of the Fast Multipole Method first developed in \cite{Greengard} to calculate gravitational forces. \Abacus{}' force-solver algorithm, introduced in \cite{Metchnik} and detailed in Pinto et al.~(in prep.), is built on the structure of the simulation space: the total simulation volume consists of a single, cubic box containing $N$ particles and an infinite number of its periodic replicas. Each box is, in turn, subdivided further into a cubic lattice of cells (with \code{CPD}$^3$ cells, where \code{CPD} stands for ``cells per dimension''). Each two-dimensional slice of the box, containing \code{CPD}$^2$ cells, is called a slab. 

To solve for the gravitational force on a single particle, \Abacus decomposes the force into a ``near field''---the force arising from nearby cells---and a ``far field''---that arising from well-separated cells.  This division is set by the \textit{near-field radius}, called $R$.  The near- and far-field terms are computed independently using separate techniques to balance accuracy with computational cost. Because the separation between the near and far fields is disjoint, every pairwise interaction is present in only one or the other.  We note that, formally, every pair of particles has an infinite number of pairwise interactions---one for each periodic image---and that distant periodic replicas of the near-field particles are included in the far-field force computation.  There is no approximation of close particle pairs as non-periodic.

The far-field acceleration is represented by Taylor-series expansions, up to order $p$, in each cell. The coefficients of these expansions are computed by cyclically convolving (in Fourier space) the multipole moments of the mass distributions in the cells with a kernel consisting of derivatives of the Green's function, which we refer to as the ``derivatives tensor''. The derivatives tensor depends only on the geometry of the grid and not on the cell's contents and may therefore be pre-computed at the beginning of the simulation and re-used at every timestep. Furthermore, both the multipole moments and the derivatives tensor possess many simplifying symmetries and recursion relationships which significantly reduce the complexity of the calculation. The multipole order parameter $p$ sets the accuracy of the Taylor approximation and the desired choice of $p$ depends on the distance between the two domains; in our case, $p=8$ yields excellent force accuracy while maintaining high performance given a separation of two cells ($R=2$).


The near-field force on a given particle is sourced by particles in neighboring cells out to radius $R$ and can be computed by any open-boundary condition Newtonian force solver; in the simulations presented, it is computed by a direct, $O(N^2)$ summation with a compact spline softening kernel. Direct summation proves to be well suited to GPUs, as the highly regular geometry of the work exposes massive parallelism and efficient memory access, communication, and computation patterns.

In both the serial and parallel implementations,  a single timestep of \Abacus{}  consists of two primary sub-steps: \code{singlestep} and \code{convolution}. We summarize the overall method in this section and describe the differences between the serial and parallel implementations of both sub-steps in the sections that follow.

\code{singlestep} flows through the grid of cells in a single direction using a ``slab pipeline'', operating on a rolling subset of contiguous slabs loaded in memory.  While the GPU is computing the near-field work, the CPU is actively evaluating the far-field force from the supplied Taylor series coefficients in that cell, packaging the near-field work for the GPU, co-adding the two forces, performing the leapfrog (kick-drift-kick) integration \citep{1997astro.ph.10043Q}, moving particles to their new cells, and computing the multipoles in these new cells.  When the slab pipeline has swept through the entire box, the global step advances and the \code{convolution} step is invoked, producing new Taylor coefficients (see below).  The near field radius $R$ sets the minimal width of the slab pipeline: the central slab requires $R$ loaded slabs on either side to compute the near force. In the event that \Abacus needs to perform on-the-fly halo finding (described further in Sec.\ref{sec:halofinding}), the pipeline must broaden sufficiently to search through slabs containing particles in the same halo.   Because the far-field force on a given cell is entirely represented by the Taylor coefficients, \Abacus never needs to load the entirety of the particle data into memory to calculate forces.  If enough memory is available, however (such as on a GPU cluster), then the full particle set may be held in memory.

During the \code{convolution} step, the Taylor coefficients are calculated that are required to evaluate the far-field force during \code{singlestep}. This is done by convolving the multipole moments of each cell computed during the previous \code{singlestep} with the pre-computed derivatives tensor to obtain the Taylor coefficients representing the approximation of the far-field potential at the center of each cell.  These may be saved to disk at the conclusion of \code{convolution}, or held in memory if desired.  The convolution can be performed in Fourier space as a multiplication, so we take the YZ-FFT during \code{singlestep} while the YZ slab is in memory. The convolution then performs the cross-slab X-FFT, multiplies it using the convolution theorem with the Fourier transform of the derivatives tensor, and concludes by applying the inverse X-FFT to obtain the Taylor coefficients. The inverse YZ-FFT is completed during \code{singlestep} immediately before applying the far-field force to a slab.

The slab pipeline makes it simple to balance CPU and GPU loads, and to mask the I/O (performed on a dedicated thread) with the bulk of the CPU work. The slab pipeline also affords \Abacus{} a unique opportunity: since only a small fraction of the simulation volume need be held in memory at any given time, \Abacus{} can run, on a single node, simulations that would force another $N$-body code to seek an allocation on a large computer cluster.  However, for \AbacusSummit, we ran entirely in-memory, since we had available the memory-rich Summit cluster.

\subsection{Parallel Method}\label{sec:ParallelAbacus}


In its single-node implementation, \Abacus' algorithm enables the execution of unprecedentedly large simulations without holding the whole state in memory.  This approach naturally lends itself to multi-node parallelization, since, to a given node, it does not matter if the not-in-memory state is on disk or resident in another node's memory.  This is how \AbacusSummit was run, with the simulation state residing entirely in memory, distributed over nodes, using the slab pipeline to organize the computation.

The parallel domain decomposition reuses the slab decomposition of the simulation volume, with each node responsible for a contiguous span of slabs at each timestep.   A given node begins a timestep with slabs $j$ through $j+N$ in memory, with slabs $j-N$ through $j-1$ on the ``left'' neighbor node, and slabs $j+N$ to $j+2N$ on the ``right'' neighbor node. The node processes its slabs sequentially, with slab $j+2$ being the first to receive forces (due to the near-field radius), and slab $k$ being the first to finish.  On a non-group-finding step, $k=j+3$, since particles from neighboring slabs must be allowed to drift into this slab, but during group-finding epochs, $k$ may be $j+10$ or greater.

Having completed all work for slab $k$, the slabs below $k$ require information from slabs below $j$, which are not in memory because they reside on the left neighbor node. The slabs below $k$, however, do not require any more information from slabs $k$ and above.  Therefore, the node can ``detach'' slabs $k-1$ and below and send them to its left neighbor node, where they will eventually be completed.  The transfer is accomplished with an asynchronous MPI send.

Because a node begins the timestep with slabs $j$ to $j+N$ in memory but ends with slabs $k$ to $k+N$ in memory, the domain decomposition is said to rotate over the nodes.  A given node is therefore not responsible for a static portion of the simulation volume, but a cyclically rotating slice.  Information is only sent once per time step and the network load is easily overlapped with other computation, since it need only arrive before the node's work has reached the other side of its domain. There is no need for nodes to frequently synchronize over stages of work, as would happen with a 2D or 3D decomposition.

In the serial implementation, at the end of \code{singlestep} all the YZ slabs' multipole moments are stored on a single node and are available to the following \code{convolution}. Subdividing the slab domain amongst the compute nodes in the parallel implementation leaves each node with a finite range in $x$ of YZ slabs' multipole moments stored in its memory. Recall that \code{singlestep} has already performed the YZ-FFT; \code{convolution} is responsible for performing the cross-slab forward and inverse X-FFTs. For this, each node requires the full range of $x$ of multipole moments slabs, but does not require the full range of $z$---the X-FFTs can be performed on chunks of arbitrary thickness in the $z$ direction. Thus, to parallelize the \code{convolution} module, we begin with an MPI \code{Alltoall} communication between the compute nodes:  each node starts with several contiguous YZ slabs of multipole moments, and ends up with several contiguous XY slabs of multipole moments held in its memory. Each node then performs the X-FFT to complete the transformation of the multipole moments to Fourier space, multiplies them using the  convolution theorem with the derivatives tensor, and takes the inverse X-FFT to obtain the Taylor coefficients. The \code{convolution} module concludes with an inverse \code{Alltoall} that stores, on each node, the Taylor coefficients belonging to the node's original YZ slab domain, leaving the nodes prepared to proceed to \code{singlestep}. 

The simplicity of the 1D domain decomposition comes at a price: the nodes need to have enough memory and enough compute power to service a slice of the box sufficient to hold the full width of the pipeline. In \Abacus{} simulations, this width is set by the need to identify dark matter halos on-the-fly, up to a natural diameter of about $10 \hMpc$. This in turn requires a computational domain of $25-30 \hMpc$ thickness, which contains about 5 billion particles. Summit's 512 GB of RAM per node enabled \Abacus{} to store a wide-enough slice to adopt this 1D parallelization scheme for the production of \AbacusSummit{}.  

The serial version of \Abacus has been extensively validated and compared to other $N$-body codes, as described in Refs.~\cite{GarrisonEtal2018}, ~\cite{GarrisonEtal2017}, and ~\cite{AbacusCode}. For given initial conditions, \Abacus generates results that are highly reproducible in the low-redshift summary statistics. At all stages of code development, we were therefore able to confirm that the parallel code base agrees to nearly machine precision in the full-simulation summary statistics, and on a particle-by-particle basis for single time steps.

\subsection{Halo Finding}\label{sec:halofinding}

\Abacus{} identifies halos with \Compaso{}: a new, hybrid, on-the-fly halo finder described in \cite{CompaSO}. Performing halo finding on-the-fly avoids the need to write and retrieve the full particle data for post-processing.  \Abacus{} uses the common friends-of-friends method \citep{1985ApJ...292..371D} to partition the particles into strictly disjoint sets. Within each set, we then use  \Compaso{} to perform a spherical overdensity (SO) method to identify science halos \citep{1994MNRAS.271..676L,2008ApJ...688..709T}. We then perform a second round of subhalo finding with a higher density contrast.  These are used as centers for halo moments, but are not reported as catalog entries in their own right.

In detail, halo finding begins by computing a kernel density estimate around all particles, using a weighting of $1-r^2/b^2$, where b is chosen to be 40\% of the interparticle spacing.  This step occurs as part of the near-field force computation with minimal computational burden, and we include the resulting density in our data products for use in other analyses. The particle set is then segmented into so-called level zero (L0) halos using the friends-of-friends (FOF) algorithm with linking length set to $b_{\rm FOF} = 0.25$ of the interparticle spacing, but including only particles with a relative overdensity contrast $\Delta = 60$. We note that the percolation density for $b_{\rm FOF} = 0.25$ is 41.8 \citep{2001JChPh.114.3659L}, well below 60.  The intention is that the bounds of the L0 halos be set by the kernel density estimate, which has lower variance than the nearest neighbor method of FOF and imposes a physical smoothing scale.  

Here and below, the density thresholds are scaled upward as the cosmology departs from Einstein-de Sitter, in keeping with spherical collapse estimates for low-density universes.  We define $\Delta$ values relative to critical density scaled by the \cite{1998ApJ...495...80B} factor of $(1 + 82x - 39x^2)/18\pi^2$, where $x=\Omega_M(z)-1$.  The FOF linking length is scaled as the inverse cube root of that change, while the kernel density scale $b$ is not changed.

All subsequent L1/L2 group finding and all halo statistics included in \AbacusSummit are based solely on the particles in their L0 halo. Within each L0 halo, \Compaso{} constructs L1 halos by a competitive spherical overdensity algorithm. The particle with the highest kernel density is selected to be the nucleus of a new halo, and \Compaso{} then searches outward to find the innermost radius in which the enclosed density of L0 particles crosses below the chosen $\Delta = 200$ threshold. Particles within this radius (labeled $R_{200}$) are tentatively assigned to the L1 group, and particles within 80\% of $R_{200}$ are marked as ineligible to be a future nucleus. Among the remaining eligible particles, \Compaso{} then locates the particle with the highest kernel density that is also denser than all other particles (eligible or not) within $40\%$ of the interparticle spacing. If the located particle has a density higher than the density generated by a singular isothermal sphere with 35 particles within $R_{200}$, the located particle is designated as a new L1 nucleus. 
For each new L1 nucleus, \Compaso{} searches in the set of all L0 particles for the $R_{200}$ SO radius. Each L0 particles is assigned to the new L1 group if it had previously been unassigned or if it is estimated to have an enclosed density with respect to the center of the new group that is twice that of the enclosed density with respect to its assigned group. The enclosed densities are estimated using an inverse square density profile scaled from the SO radius. 
We note that \Compaso{} does not perform any  unbinding of particles based on their gravitational potential, as in the case of other halo finders, e.g.~\textsc{ROCKSTAR} \citep{Behroozi_2012}.

Within each L1 halo and using only the L1 particles, \Compaso{} repeats the SO search to locate L2 halos with an enclosed density contrast of 800. \AbacusSummit uses the center of mass of the largest L2 subhalo to define a center for computing L1 halo statistics.  \AbacusSummit further outputs the masses of the five largest L2 subhalos; however, no further information about the L2 substructure is included in the \AbacusSummit data products.

\Compaso{} has been compared extensively and compared against existing halo finders; the results are described in \cite{CompaSO}.  

\subsection{Merger trees and cleaned halo catalogs}

Following the identification of spherical overdensity halos using the \Compaso{} algorithm, we construct halo merger trees as a post-processing step using the 33 halo output epochs. Halo merger trees correspond to associations between L1 halos identified across output times, resulting in lists of progenitor and descendant halos. Our merger tree algorithm works in a {\it reverse time order}, i.e., traversing the halo catalogs from low to high redshift.

To accelerate the process of associating a halo, {\tt halo\_now}, at snapshot {\tt i}, with its progenitors in snapshot {\tt i-1}, we ``pre-filter'' the \Compaso{} catalog by first selecting only those halos at snapshot {\tt i-1} that are within at most a distance of 4 Mpc$/h$ from {\tt halo\_now}. This narrows down the search to a much smaller list of plausible halo associations. 

From this list, we then identify {\it candidate associations} as those halos that donate a non-zero fraction of their unique particle IDs, $f_{{\rm donate}}$, to {\tt halo\_now}. We also record the fraction of subsampled particles in {\tt halo\_now} that are donated to it by its candidate associations as $f_{{\rm match}}$. Note that all matched fractions are weighted by the particle kernel density, which gives preferences to associations that donate particles to the central core of {\tt halo\_now}. Finally, we mark a candidate association as a {\tt Progenitor} if $f_{{\rm donate}}\geq0.5$. Furthermore, the {\tt MainProgenitor} is identified as the association that contributes the largest $f_{{\rm match}}$. We repeat this procedure for all \AbacusSummit halos containing at least 50 particles, and with at least 5 subsample particles available for tracking. 

A primary application of our merger trees is to post-process and ``clean'' the raw \Compaso{} halo catalogs output by \Abacus{}. This procedure involves identifying halos with non-monotonic mass growth as a result of one or more dynamical events in their past histories including fly-bys, halo ``splits'' (where a single halo may be deblended by \Compaso{}), partial mergers etc. We identify these objects as those whose peak mass is at least two times greater than their present day mass. We then traverse the merger tree of each flagged object to find a nearby, neighboring halo that shares the same progenitors as the flagged halo; the two halos are then merged at that redshift and remain merged at all subsequent times. The net effect is to ``clean'' the \Compaso{} catalogs of several low mass halos, typically found at the peripheries of larger systems. The masses of these larger halos are then incremented by the mass of the individual halos merged on to them. The full details of the cleaning algorithm are described in \cite{MergerTree}, where we also demonstrate the various benefits of cleaning the \Compaso{} catalogs using this method. 
\subsection{Light Cones}\label{sec:lightcones}

\AbacusSummit simulations produce a light cone centered at the corner of the box and include one periodic copy of the box displaced in the $y$ direction, and one displaced in the $z$ direction. At every timestep, \Abacus identifies particles that belong to both the light cone and the 10\% particle subsample (the concatenation of the A and B subsamples; see Section~\ref{sec:DataProducts}), and outputs their positions, velocities, particle IDs, and HEALPix pixel number that can be used to form projected density maps. The HEALPix pixels are computed using +z as the North Pole, i.e., the usual (x,y,z) coordinate system, with the Nested pixel labeling.

For base boxes with length 2000$\hMpc$ on a side, we position the light cone observer at (--990, --990, --990), or, in other words, 10 $\hMpc$ inside a box corner. \Abacus sweeps through the box and two of its periodic copies---these three boxes forming the eligible space of the light cone are centered at (0,0,0), (0,0,2000), and (0,2000,0), respectively (where all lengths are given in $\hMpc$ units). This provides an octant to a distance of 1990 $\hMpc$ ($z=0.8$), shrinking to two patches each about 900 square degrees at a distance of 3990$\hMpc$ ($z=2.45$).

The three copies of the box are output separately and the particle positions are referred to the center of each periodic copy: e.g.~the particles from the higher redshift box need to have 2000 $\hMpc$ added to their z coordinate.  

For the two \code{huge} boxes of $7.5\,h^{-1}\mathrm{Gpc}$, we position the light cone observer at the center of the box (0,0,0).  This provides a full-scale light cone to 3750\hMpc\ distance ($z=2.18$), with smaller areas reaching to 6495\hMpc\ in the 8 corners.  For example, half the sphere is available to 4500\hMpc\ ($z=3.2$).

The algorithm for identifying particles belonging to the light cone is as follows. At each timestep, \Abacus computes the radii of the light cone at the beginning and at the end of the step. The goal is to identify every particle which, during the upcoming timestep, will pass through the light cone. To do this, we check each cell in the simulation box to determine whether its center is in the vicinity of the light cone. If not, the cell may be skipped. If the cell is close to the lightcone, however, \Abacus ``opens'' the cells and considers each particle in the given cell individually. 

Using velocity-extrapolated leapfrog integration, \Abacus computes the fraction of the timestep (\code{fracstep}) when a given particle is expected to meet the lightcone sweeping through its host cell (accounting for periodic boundary wraps as needed). If \code{fracstep} is between zero and one, this implies that the particle will cross the lightcone during the given timestep and therefore should be flagged for output.  The particle's position and velocity are extrapolated using drifts and kicks, respectively, to the time at which it will cross the light cone. To guard against floating point errors arising from subtracting two numbers close in magnitude when computing \code{fracstep}, we accept values of \code{fracstep} just below zero (in the case of \AbacusSummit, we set the tolerance to $10^{-6}$), so that particles that fall just behind the lightcone during the timestep are nevertheless include in the output.

We note that we do not perform extrapolation across the transverse boundaries of the light cone (e.g. the boundaries bordering the edge of one of the three periodic copies of the box that is not adjacent to another copy). Therefore, the light cones are accurate only out to a distance corresponding, conservatively, to the width of one \Abacus cell (\code{BoxSize}$/$\code{CPD}) from their transverse edges. In the case of most \AbacusSummit simulations, this distance is equal to approximately $1.2$ $\hMpc$. However, this estimate is conservative; a particle typically moves much less than the width of a cell per timestep, so it is therefore highly unlikely that a particle $1.2 \hMpc$ away from the transverse edge would move sufficiently far such that the lack of the periodic wrap would fail to account for it.




\section{Performance on Summit}\label{sec:Performance}
\subsection{Overview}
\Abacus on Summit is very fast. We achieve roughly 70M particles/sec per node at early times, until $z\sim 1$.  Past that time, the computation becomes dominated by the near-force calculation on the GPUs, with the performance falling to 45M particles/sec per node at $z=0.1$.  For comparison, a high-performing example of a previously published $N$-body speed was 3.8M particle-steps/sec/node on Titan with \textsc{pkdgrav3} \citep{2017ComAC...4....2P}, although this result was on older hardware and we expect improved results could be presented by those authors.

\Abacus owes its high performance to its novel force solver, which enables \Abacus to run using the slab pipeline structure described in Section~\ref{sec:Abacus}, dispatching significant work to both the CPU and GPU and overlapping their computation. In addition, we tune for performance with hardware-specific optimizations. In this section, we describe Summit's system and hardware, the achieved performance of \Abacus on Summit, the optimizations and features that contribute to \Abacus' speed in general and on Summit specifically, and, in brief, the performance of ancillary code modules used in the \AbacusSummit production pipeline. 


\begin{deluxetable*}{l|r|r|D|r|r|r|r|r}[htp]
\centering
\tablehead{ \colhead{Spec.}  &
\colhead{$N$}  &
\colhead{Size} &
\twocolhead{Final $z$}  &
\colhead{Num.} &
\colhead{Nodes} &
\colhead{Total node-hours} &
\colhead{Wall-clock,} &
\colhead{Data products,} \\[-2ex]
\colhead{} & \colhead{} & \colhead{[Mpc/$h$]} & \colhead{} & \colhead{} & \colhead{boxes} & \colhead{} & 
\colhead{(per sim)} &
\colhead{per sim} &
\colhead{compressed}
}
\decimals
\startdata
\texttt{base}      & $6912^3$ & 2000 & 0.1 &  139 &  60 & 250 K (1800) & 30 hr & 1600 TB \\
\texttt{huge}      & $8640^3$ & 7500 & 0.1 &    2 & 141 & 6.5 K (3300) & 23 hr & 100 TB \\
\texttt{high}      & $6300^3$ & 1000 & 0.1 &    1 &  46 & 1.9 K (1900) & 41 hr &  40 TB \\
\texttt{hugebase}  & $2304^3$ & 2000 & 0.1 &   25 &   5 & 1.1 K (43)   & 8.6 hr & 22 TB \\
\texttt{highbase}  & $3456^3$ & 1000 & 0.1 &    2/1 &  15/35 & 790 (260) & 16/8.7 hr & 12 TB\\
\texttt{fixedbase} & $4096^3$ & 1185 & 0.1 &    5/1 &  30/15 & 2.6 K (440) & 15/14 hr & 39 TB \\
\texttt{small}     & $1728^3$ &  500 & 0.2 & 1643 & 1 & 32 K (20) & 20 hr & 140 TB \\
\hline
Total & 57 T & $2400\,(\mathrm{Gpc}/h)^3$ & ~ & ~ & ~ & 290 K & ~ & 2000 TB \\
\enddata
\caption{Computational resources used in the production of \AbacusSummit.  \texttt{highbase} used 15 nodes for two sims, and 35 for one. \texttt{fixedbase} used 30 nodes for five sims, and 15 for one. Of the \texttt{small} boxes, 1643 reached the final redshift of $z=0.2$; 1883 reached $z=1.1$. Only the former are presented in this table to aid interpretation of the timings. \label{tbl:node_hours}}
\end{deluxetable*}

\subsection{The Summit System}

Summit is comprised of 4608 IBM AC922 compute nodes, each with two 22-core IBM POWER9 processors and six NVIDIA Tesla V100 GPUs. Each compute node contains 512 GB of RAM and 96 GB of High Bandwidth Memory accessible by the GPU accelerators. The compute nodes each provide a theoretical double-precision capability of 40 TF \citep{summitUserGuide}.

Each IBM POWER9 processor utilizes IBM's SIMD Multi-Core technology (SMC); SMCs support simultaneous multi-threading (SMT) up to level 4, such that each physical core is capable of running a maximum of 4 hardware threads (compare with Intel's hyper-threading). The POWER9 processor contains 22 SMCs. To maximize our performance on the POWER9 processors, we have added VSX intrinsic functionality to relevant code modules within \Abacus (Section \ref{sec:vsx}), augmenting our existing Intel AVX capability.

The Summit node-hour usage is presented in Table~\ref{tbl:node_hours}, organized by simulation specification (\texttt{base}, \texttt{hugebase}, etc.; see Sec.~\ref{sec:sims}).

\subsection{Performance}

\begin{deluxetable*}{l|D|r|D|r}[htp]
\centering
\tablehead{ \colhead{}  & \twocolhead{High-z}  & \colhead{} &  \twocolhead{Low-z}  & \colhead{} \\
\colhead{Action} & \twocolhead{Time [s]} & Rate & \twocolhead{Time [s]} & Rate}
\decimals
\startdata
Total & 79.5 & 69.2 Mp/s & 122.0 & 45.1 Mp/s \\
\hline
\quad Convolution & 11.7 & \ & 11.8 & \ \\
\qquad Fourier Transforms & 3.6 & \ & 3.6 & \ \\
\qquad Convolution Arithmetic & 5.0 & \ & 5.0 & \ \\
\qquad Array Swizzle & 2.1 & \ & 2.1 & \ \\
\qquad Disk I/O & 0.9 & \ & 1.1 & \ \\
\hline
\quad Singlestep & 66.6 & 82.6 Mp/s & 108.8 & 50.6 Mp/s \\
\qquad CPU Work & 63.1 & \ & 58.8 & \ \\
\qquad \quad Transpose Positions & 4.4 & \ & 4.6 & \ \\
\qquad \quad Index Near Force Pencils & 2.4 & \ & 2.4 & \ \\
\qquad \quad Taylor Force & 17.0 & 324 Mp/s & 15.6 & 353 Mp/s \\
\qquad \quad Kick & 4.6 & \ & 3.9 & \ \\
\qquad \quad Drift & 3.2 & \ & 2.9 & \ \\
\qquad \quad Finish & 22.7 & \ & 22.6 & \ \\
\qquad \quad \quad Merge New Particles & 2.4 & \ & 2.3 & \ \\
\qquad \quad \quad Compute Multipoles & 13.7 & 401 Mp/s & 12.9 & 427 Mp/s \\
\qquad \quad *Group Finding & . & \ & 193.1 & 28.5 Mp/s \\
\qquad \quad Unaccounted & 8.8 & \ & 6.8 & \  \\
\qquad GPU Near Force (non-blocking) & 39.4 & 140 Mp/s & 96.7 & 56.9 Mp/s\\
\qquad Waiting for GPU, MPI, or I/O & 2.4 & \ & 44.9 & \ \\
\enddata
\caption{Wall clock timing for a typical \AbacusSummit{} time step, from the  \code{AbacusSummit\_base\_c000\_ph000} box realization ($N_P = 6912^3$, \code{BoxSize} = $2$\hGpc) executed on 60 compute nodes. Timings are shown for two representative steps, one at $z=90.5$ and the second at $z=0.105$  (\textit{Group Finding} from $z=0.1$ also shown). Units of Mp/s denote millions of particles per second.    \textit{Non-blocking} means other CPU actions can proceed while that action is running.  \textit{Unaccounted} time is time spent checking preconditions, or other asynchronous overheads. \label{tbl:timing}}
\end{deluxetable*}

\begin{figure}
    \centering
    \includegraphics[width=\columnwidth]{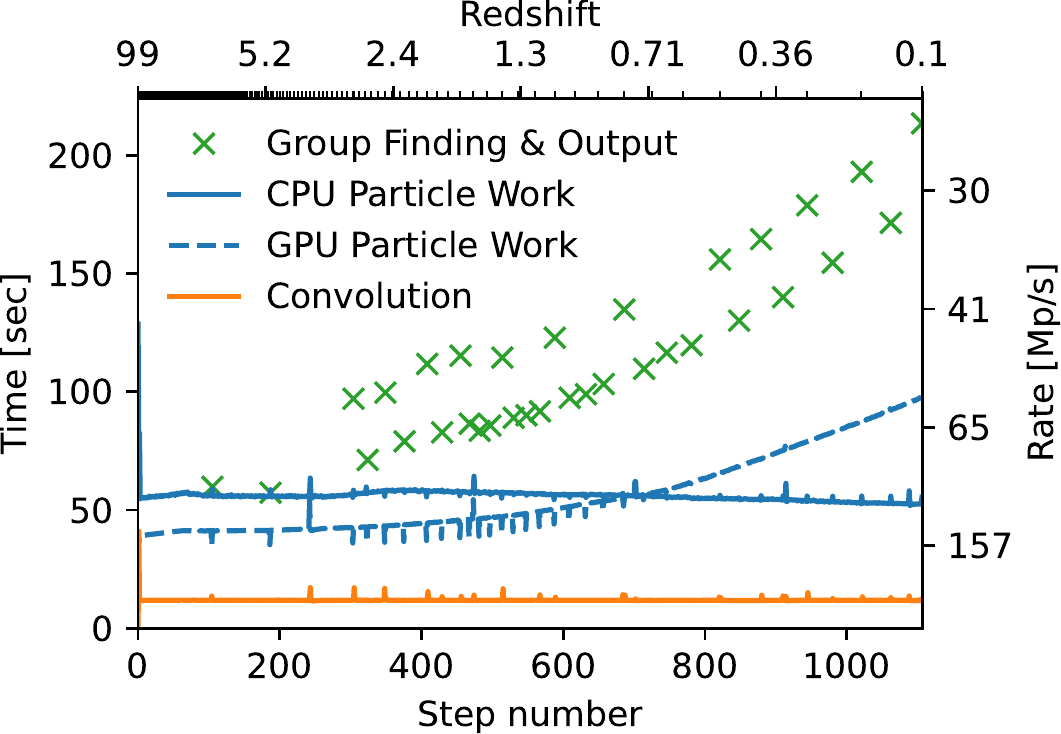}
    \caption{A timing overview of a single simulation (\code{base\_c000\_ph000}).  CPU work (solid blue line) includes Kick, Drift, Multipoles, etc. (Table \ref{tbl:timing}) and is approximately constant; GPU near-field force time (dashed blue line) increases as the clustering in the simulation increases.  The CPU and GPU work run concurrently.  The Convolution (solid orange line) precedes the CPU/GPU work at each time step.  Group-finding occurs at 33 epochs, 12 of which include full particle outputs.}
    \label{fig:timing_overiew}
\end{figure}

\begin{figure}
    \centering
    \includegraphics[width=\columnwidth]{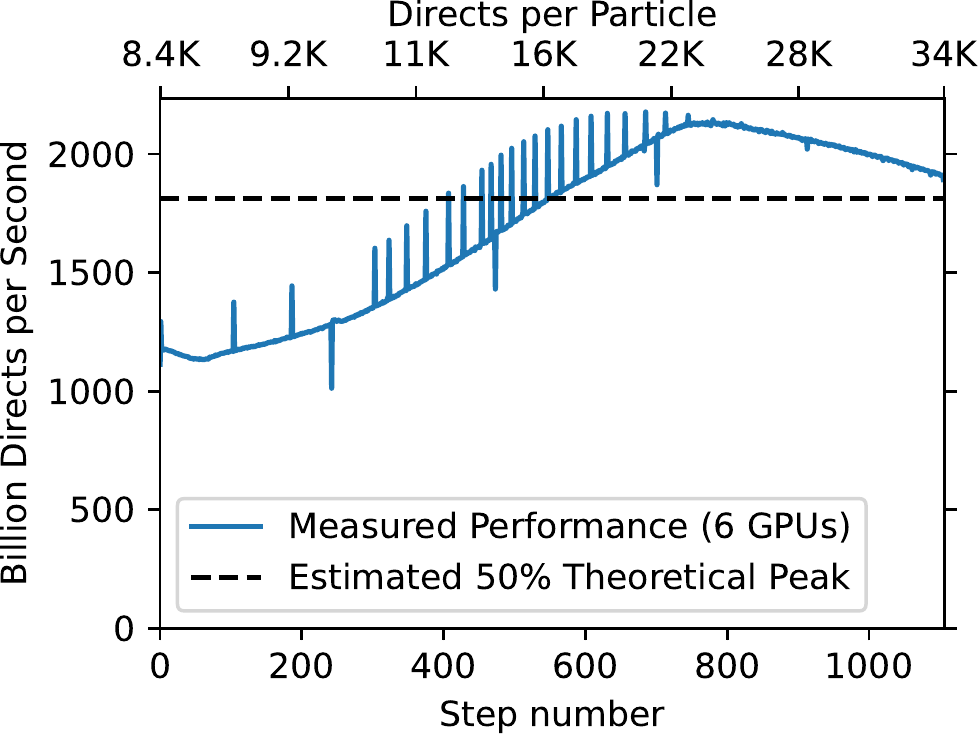}
    \caption{The GPU performance (mean across nodes) for the \code{base\_c000\_ph000} simulation, in billions of pairwise interactions (directs) per second, including overheads like host-device communication.  The theoretical maximum rate is computed assuming 26 floating point operations per interaction (see text), and assuming all of those operations can be executed with FMA---a highly conservative approximation.  Under these assumptions, we observe a peak of 60\% of theoretical performance.}
    \label{fig:gpu_performance}
\end{figure}


The main \Abacus{} code consists of two executables invoked in a tick-tock fashion: \code{singlestep} and \code{convolution}.  \code{singlestep} dominates the runtime---it computes forces and performs the particle update cycle, while \code{convolution} only operates on the cell multipoles.  Table~\ref{tbl:timing} presents timings for these two executables for a typical \AbacusSummit{} box ($N_P = 6912^3$, \code{BoxSize} = $2$\hGpc) executed on 60 compute nodes.  Timings at two epochs are shown: $z=90.5$ (nearly unclustered), and $z=0.105$ (highly clustered).  The total step time is 79.5 seconds (69.2 million particles per second) at the former epoch, and 122 seconds (45.1 million particles per second) at the latter.  Figure~\ref{fig:timing_overiew} shows the evolution of the runtime throughout the course of the simulation.

\code{convolution} takes 10--15\% of each step's runtime.  The workload of \code{convolution} is independent of epoch, consistently requiring approximately 12 seconds for a typical \AbacusSummit box.  Of this time, about a third of the work occurs in executing fast Fourier transforms before performing the arithmetic required to apply the Convolution Theorem, which in turn accounts for about $40\%$ of \code{convolution}'s runtime. An additional 2 seconds are required to re-order (swizzle) arrays before and after taking their transforms, with an additional 1 second to read the lightweight Derivatives files.  The GPUs are not used in this part of the code.

\code{singlestep} accounts for 85--90\% of an \Abacus{} timestep's runtime. In this portion of the code, \Abacus{} sweeps through the particles slab-wise, executing a series of pipeline actions on each slab.  The CPU loads the particle positions, constructs the indexing structures for the GPU near-field work, computes the far-field force by evaluating the Taylor series of the potential, kicks the particle velocities, drifts the positions, and reorders particles into their new cells.  Finally, the cell multipoles of the new particle positions are computed.  The GPU computes near-field forces while the CPU is executing the rest of this work.

As described in Sec.~\ref{sec:AbacusSummit}, we choose the code parameter \code{CPD} (cells per dimension) such that the CPU and GPU workloads are balanced.  Since the GPU work grows as the simulation becomes more clustered, we cannot balance the work for all time steps, but we can choose a value that minimizes the time-to-solution.  For the fiducial $N_P = 6912^3$ \AbacusSummit{} simulations, this optimum is around $\texttt{CPD}=1701$, or 67 particles per cell.

With this \texttt{CPD}, the CPU work masks the GPU computation prior to $z\sim 1$.  In this early regime, \code{singlestep} takes about 67 seconds (83 million particles per second) and the CPU work accounts for 63 of those, with the remainder being attributed to fixed startup and teardown costs (most from MPI, CUDA, and load balancing).  The GPU near-field work takes only 39 seconds and is thus completely masked.

After $z\sim1$, the near-field work dominates the runtime, reaching 97 seconds by the final redshift $z=0.1$.  \code{singlestep} takes 109 seconds in total at this epoch; the 12 second differential arises because the final kick, drift, MPI send/receive, merge, and multipoles cannot happen until the final GPU forces are computed, in addition to the usual fixed startup and teardown costs.

Considering the \code{singlestep} CPU work (59--63 seconds), almost half of the time is spent in two stages: the Taylor Force (15--17 seconds) and the Multipoles (13--14 seconds).  The former evaluates the gradient of the Taylor series of the potential, producing the far-field force, and the latter evaluates the cell-wise multipoles for use in the convolution.  Both stages include a 2D (slab-wise) FFT.  These stages contain most of the floating-point CPU work and benefit from SIMD acceleration.  The remaining half of the CPU time is spent in memory-bound work, like the Kick, Drift, Transpose, and Merge.

On Summit, \Abacus has achieved a top GPU performance of 2200G direct pairwise force calculations per second (Figure~\ref{fig:gpu_performance}), including host-GPU communication, notionally about 56 TFlops of compute speed and 60\% of peak theoretical performance.  The peak theoretical performance is computed by scoring each arithmetic operation in the direct kernel as 1 FLOP, and the inverse square root as 2 FLOPs, following the guidance of the NVIDIA Nsight documentation\footnote{\url{https://docs.nvidia.com/gameworks/content/developertools/desktop/analysis/report/cudaexperiments/kernellevel/achievedflops.htm}}.  This yields 26 FLOPs per interaction.  The six V100 GPUs are assumed to be running constantly at their boost-clock performance of 15.7 TFLOPS\footnote{\url{https://images.nvidia.com/content/technologies/volta/pdf/volta-v100-datasheet-update-us-1165301-r5.pdf}}, which also assumes every operation is a 2-flop fused multiply-add (FMA).  Since not all of the floating-point operations in our kernel can be expressed as FMA, we are likely close to the peak performance that can be achieved for this particular kernel implementation.  The lower off-peak performance can be attributed to lower mean flop-to-byte ratio at early times, and the large number of sparse (void) kernels at late times.

The GPU throughput could have been modestly improved in several ways, but at a worse time-to-solution.  Larger cells (lower \code{CPD}) would increase the FLOP density, but require more FLOPs overall.  We additionally could have more efficiently overlapped GPU compute and communication by using three CUDA streams per GPU instead of two, but this negatively impacted CPU performance and resulted in a slower runtime.  In early-time, I/O-bound steps where the CPU spent a large fraction of time idle, the GPU performance increased about 20\%, which is seen as the spikes of increased performance in Fig.~\ref{fig:gpu_performance}.  This indicates that host-side resource contention is likely an early-time bottleneck, and with less memory pressure during the I/O steps, the positions and accelerations can be filled and communicated more efficiently.

In \code{singlestep}, network loads are low; there is only one burst of transfer along the 1-d torus described in Sec.~\ref{sec:Abacus}, and it takes about 5\% of the runtime, arriving well before it is needed. Throughout the production of \AbacusSummit, we consistently saw an additional several seconds of MPI-related spinning in our timesteps. We speculate this spinning occurred as a result of delays in the compute nodes accessing memory buffers allocated internally by the MPI library. Additionally, operating system functions such as \code{munmap()} contributed to the runtime, albeit mostly in a non-blocking manner.  In the case of \code{munmap()}, we devote a separate thread to freeing POSIX shared memory so that it is able to run in the background while the CPU work proceeds.  It requires on the order of 25 seconds and is non-blocking, but likely contributes extra memory pressure \citep[for more on the usage of POSIX shared memory and the associated overheads, see][]{2021arXiv210213140G}.

The outer-most level of \Abacus{} consists of a Python driver script, which loops over the timesteps and invokes the parallel job dispatcher (called \code{jsrun} on Summit, which uses IBM's Spectrum LSF batch scheduling system) once per timestep.  In addition, the Python driver periodically checkpoints the simulation state by launching a copy from each node's POSIX shared memory to network storage.  A typical checkpoint takes on the order of 110-150 seconds and runs 1-2 dozen times, depending on the simulation specifications and our optimism about the cluster state.  Finally, while each \code{jsrun} call incurs an overhead of less than 1 second, there is some overhead to relaunching the executable and initializing libraries like CUDA and MPI for each time step.  Future updates to \Abacus{} will obviate the need to invoke a \code{jsrun} command multiple times per simulation.

The \code{base} simulation boxes required about 1100 time steps to reach $z=0.1$ from an initial redshift of $z_{\textrm{init}} = 99$. A typical \code{base} simulation---including the Python wrapper, backups, and main \Abacus execution---required 1800 node-hours to complete.

\subsection{SIMD Multipoles and Taylors}\label{sec:vsx}

We optimize the CPU computation of the multipoles and Taylors using single-instruction, multiple-data (SIMD) instructions (also known as vector instructions). We use the vector extensions of the POWER9 AltiVec instruction set via the built-in VSX ``intrinsic'' functions of the GCC compiler.  As described in the Abacus code paper \citep{AbacusCode}, we use a Python meta-code to manually unroll the triple-nested loop over multipole order that the multipole and Taylor computations require.  The meta-code emits intrinsics to process two double-precision particle coordinates per 128-bit VSX vector, interleaving multiple VSX vectors to mask instruction latency.

We find that an eight-fold unrolling of the particle loop (i.e.~interleaving 4 VSX vectors) is optimal.  A POWER9 core is capable of launching 2 VSX vector operations per cycle, for a peak of 8 double-precision operations per cycle, assuming FMA\footnote{\url{https://openpowerfoundation.org/?resource_lib=POWER9-processor-users-manual}}.  With the interleaved code, we find that SMT1, 2, and 4 perform the same, while SMT 2 and 4 perform better without interleaving.  This is consistent with the understanding that SMT allows the processor to mask latency by ``backfilling'' with instructions from other hardware threads, but that this is not necessary if the interleaving already masks the latency.

Although the VSX vectors are relatively narrow, each Summit node has 42 user-facing cores per node, distributed over 2 sockets (1 core per socket is reserved for system use).  We use 35 of these for floating-point work (or, more specifically, 70 hardware threads because other areas of the code benefit from SMT2).  In production simulations, we process the multipoles kernel at about 740 million particles per second, or about 20 per core.  Artificial benchmarks were only a few to 10 percent faster per core.

The Taylors kernel computation is slower per-particle than the multipoles, but is more efficient in terms of floating-point operations per second (FLOPS), because the FLOP count per-particle for the Taylors is several times that of the multipoles. The VSX calculations all give the same result as their non-vectorized counterparts to within rounding error.

\subsection{Thread Scheduling and NUMA}
Summit nodes employ a non-uniform memory access (NUMA) architecture, with each of its two POWER9 processors associated with one half of the system memory.  Memory access within the associated half (within the NUMA node) is faster than access to the other half.  Our NUMA strategy is to divide each slab between the two NUMA nodes.  Within our OpenMP thread pool, each thread only works on the half of the slab belonging to its core's NUMA node.  Threads are pinned to cores with the OpenMP affinity mechanism, which also ensures no costly migration of threads between cores or contention between threads.  We employ a custom OpenMP scheduler, written using OpenMP Tasks and atomic addition, which allows dynamic scheduling of threads within their NUMA nodes.

For the production of \AbacusSummit, we use the POWER9's simultaneous multi-threading (SMT) to use multiple hardware threads per CPU core.  This improves performance of some areas of the code by allowing each core to run several instruction streams simultaneously.  Of the three available SMT levels---SMT1, SMT2, SMT4---we have found that SMT2 (two hardware threads per core) produces the best performance.  SMT2 yields 84 available threads---we use 70 of those (35 physical cores) for OpenMP threads, 12 threads dedicated to interfacing with the GPUs, and 1 thread dedicated to running \code{munmap()} to free pages of memory, and 1 thread working on I/O.

In the computation of the near-field force, each slab is divided into spatially compact GPU work units.  Work units are dispatched to GPUs via queue system, with one queue per NUMA node.  Work units are dispatched to the queue of their NUMA node.  Each of the 12 GPU threads (2 per GPU) listens to the queue of its NUMA node, hence work is dynamically dispatched to GPUs while maintaining NUMA locality.  Host-GPU communication is done via pinned memory buffers; the initial pinning time at the beginning of each time step is a noticeable overhead but is masked by the convolution.

\subsection{Production Pipeline}
With over 150 simulation boxes to run, automation of the steps to prepare, run, and post-process each simulation was a priority, so as to increase efficiency of production and avoid human error.  These tasks were assembled in a Python pipeline that generated the simulation parameter files, queued the generation of initial conditions on OLCF's Rhea system, executed \Abacus on Summit, queued post-processing on Rhea, and finally sent the data products to OLCF HPSS (tape) and to NERSC.  The pipeline was organized as a progression of ``stages'', each with an ``indicator'' that detected whether a particular stage had completed by examining the queue state or reading files on disk, for example.  The current stage of a simulation box was compared against its last-known stage, to check for regressions or an inconsistent state.  This production pipeline handled the flow of the simulations across the four systems involved in the creation of \AbacusSummit: Summit, Rhea, OLCF HPSS, and NERSC.

While the execution of \Abacus is by far the lengthiest step in the production pipeline, both the generation of initial conditions and the post-processing constitute important parts of the overall suite production. We summarize them below.

\subsection{Initial Conditions}
\Abacus{} uses \code{zeldovich-PLT} \citep{GarrisonEtal2016} to generate initial conditions on ORNL's \code{rhea} computer. \code{zeldovich-PLT} is designed to run out-of-core, buffering state on disk, and is therefore capable of producing initial conditions on a single node even for very large problems.  Given the availability of the Rhea cluster at OLCF and the fast Alpine network file system, we opted to use \code{zeldovich-PLT} out-of-core, rather than parallelize it for distributed memory systems.  The performance of the code was bound by the I/O speed, but the runtime (18 node-hours) was still small compared to the time to execute a simulation, and so did not need to be optimized further.

\subsection{Post-processing}
During the course of a simulation, \Abacus writes most data products in a raw binary format.  Each simulation slab is written to a separate file, so the I/O is trivially parallelized over ranks.  These raw binary formats are efficient to write, but are not optimized for space and are not self-documenting.  Therefore, after reaching its terminal redshift, each \AbacusSummit box is post-processed to compress the final data products and package them in a self-describing file format (ASDF, Section \ref{sec:file_formats}) before being transferred to NERSC and OLCF HPSS.

The set of post-processing tasks is quite heterogeneous, with large variations in the data volumes across redshift and, in the case of light cones, across nodes.  The number of tasks could be 10s of thousands---too many to send to the batch scheduler, with a large uncertainty in the proper amount of time to request for each, especially given the reliance on network file system performance.  Therefore, to mitigate the heterogeneity, we use the \textsc{disBatch}\footnote{\url{https://github.com/flatironinstitute/disBatch}} code to do dynamic dispatch within a given resource allocation.  The individual tasks are itemized in a ``task file'', with one command-line invocation per line, and the \textsc{disBatch} engine sends jobs to each node in a batch allocation until the nodes are filled.  New tasks are dispatched as jobs finish and resources become available.  The return code of each task is collected by the engine, and the results are recorded in a status file that can be used to retry failed jobs.

The post-processing performance depends on the disk speed of the OLCF Alpine file system, but a typical post-processing job takes from a few to a dozen node-hours, depending on whether or not the simulation box outputs full time slices.

\section{Cosmological Opportunities with \AbacusSummit{}}\label{sec:Results}

\subsection{Scope}

\begin{figure}[tb]
\centering
    \includegraphics[width=\columnwidth]{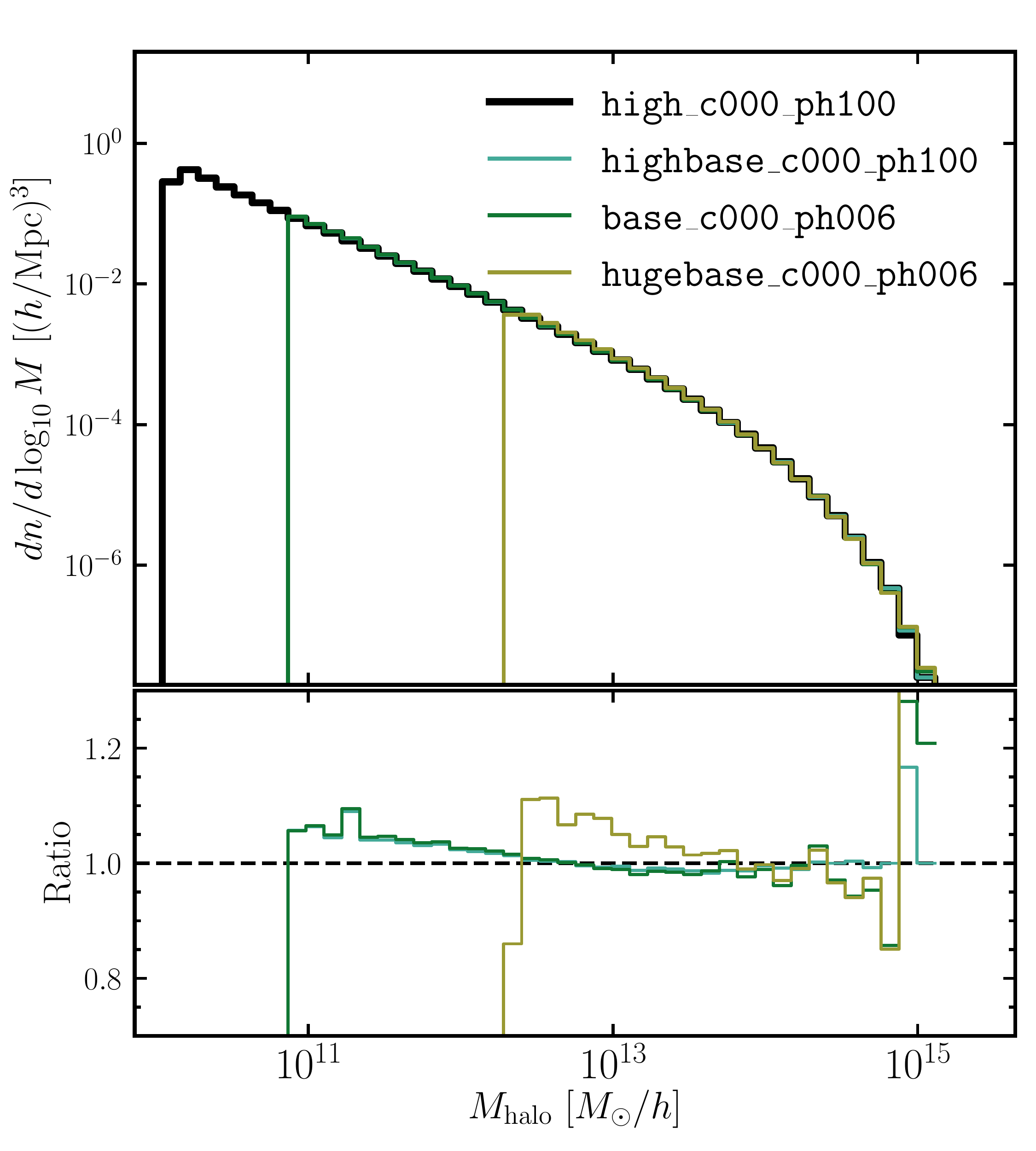}
    \caption{The halo mass function at $z = 0.5$ using the cleaned catalogs from four different simulations and three different mass resolutions, all for the \code{c000} cosmology.  We present two phase-matched pairings: 
    \code{highbase} compared with \code{high}, and then \code{hugebase} compared with \code{base}. 
    The bottom panel shows the ratio of these curves with respect to the \code{high} resolution box.  One sees that the two simulations with the same mass resolution, \code{base} and \code{highbase}, are in excellent agreement.  When comparing a coarser simulation to a finer one, there is a mild excess of halos at sizes below a few hundred particles.
}\label{fig:massfunction}
\end{figure}


We have designed \AbacusSummit{} to be superb at identifying halos in the context of high-precision large-scale structure, with the intention of enabling wide-ranging studies of the clustering of galaxies and matter as a function of cosmological parameters.  

The data volume and diversity is sufficiently large that it is difficult to summarize.  One metric is the number of L1 halos catalogued.  Summing over all simulations to our catalog limit of 35 particles, this totals $56.0$ billion at $z=2$, $68.4$ billion at $z=1.1$, and $67.3$ billion at $z=0.2$ (a slight decrease due to the smaller number of covariance boxes that reached $z=0.2$).
Restricting to halos with more than 250 particles, about $5\times 10^{10}\hMsun$ in most simulations, we catalog $5.7$ billion at $z=2$, $9.1$ billion at $z=1.1$, and $10.7$ billion at $z=0.2$.

Figure \ref{fig:massfunction} shows the mass function of halos in each of our 3 mass resolutions, using the cleaned halo catalogs and the \code{c000} cosmology.  One sees the obvious sign of the mass cutoff, but also a mild excess in the number of halos at a given mass for halo sizes up to a few hundred particles.  We note that this may be interpreted as the mass of a halo at coarse resolution being mildly higher than the same halo at the finer resolution, by roughly 5\%.  Such dependence on the number of particles is not uncommon in halo finders and could result from a number of causes.  Finer resolution gives more ability to detect and deblend neighbors or a better centering of the primary halo.  It may also be that coarser resolution creates noise in the mass estimate, which given the steeply falling mass function tends to produce a small bias in the inferred mass.  We conclude that applications depending on halo mass need to account for variations in definitions between different halo finding methods and resolutions.

\begin{figure}[tb]
    \centering
    \includegraphics[width=\columnwidth]{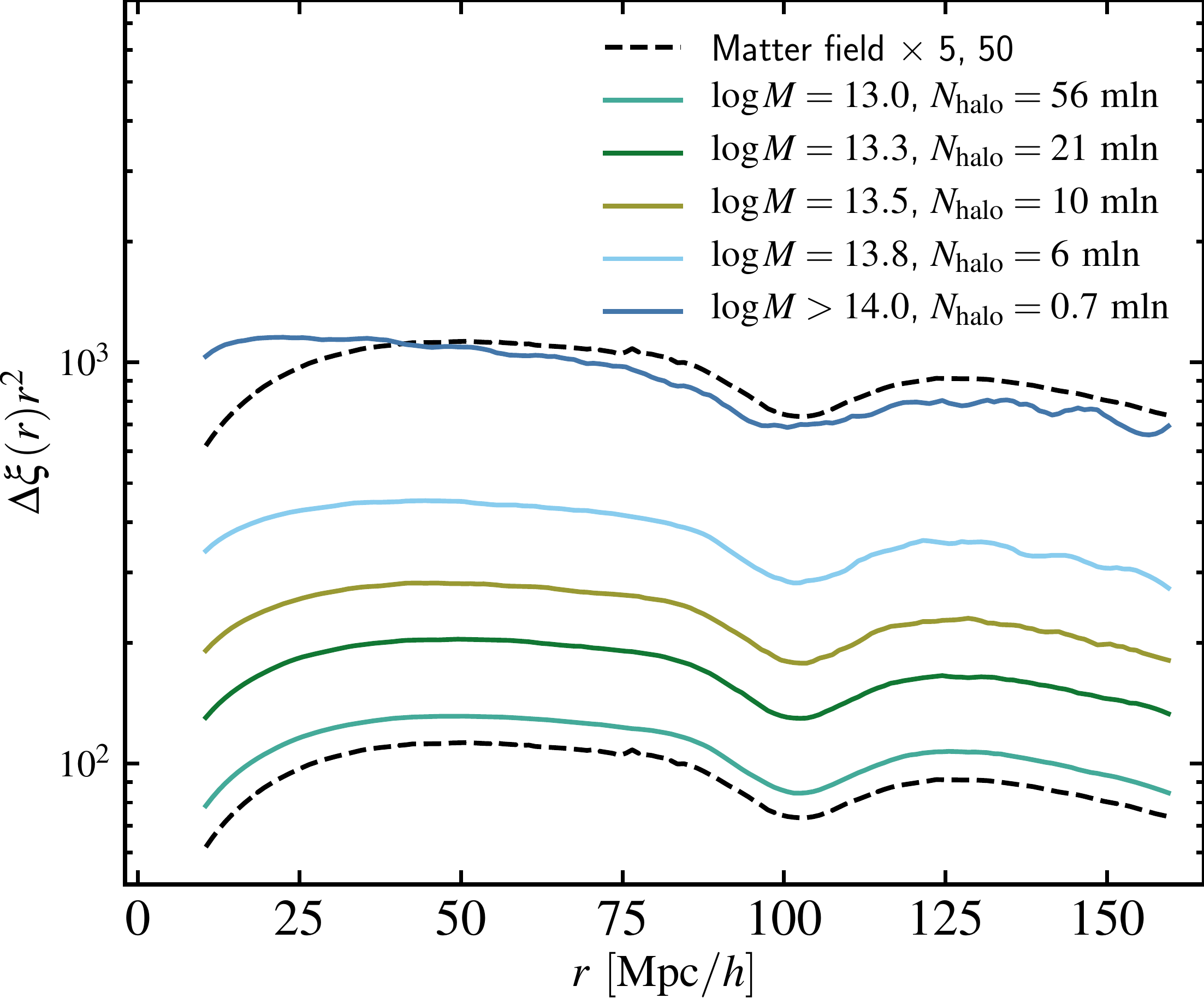}
    \caption{Modified correlation function of the halos, $\Delta \xi(r) \equiv \bar\xi(<r) - \xi(r)$, for the $7.5\hGpc$ huge box simulation at $z = 1.1$ as a function of scale, $r$. We split the halos into 5 mass bins in $\log M$: [12.9, 13.1], [13.2, 13.4], [13.4, 13.6], [13.6, 14] and $>$ 14. The average halo mass in the heaviest bin is $\log M \approx 14.15$ in units of $M_\odot/h$. The black dashed curves show the modified correlation function $\Delta \xi(r)$ computed for the matter field of the particles in subsample \texttt{A} and \texttt{B} and multiplied by a factor of 5 and 50 to aid the visual comparison.}
    \label{fig:corrfunc}
\end{figure}

Figure \ref{fig:corrfunc} shows the large-scale correlation function of halos in one of the $7.5\hGpc$ huge box simulations.  We use a variety of mass thresholds, starting at 200 particles, and compute the real-space two-point correlation function $\xi(r)$ of halo centers at $z=1.1$. For pedagogical purposes, we show a modification
\begin{equation}
    \Delta\xi(r) = \bar\xi(<r) - \xi(r)
\end{equation}
where $\bar\xi(<r)$ is the average of $\xi$ inside a sphere of radius $r$: $(3/r^3)\int_0^r \xi(r') r'^2 dr'$.  We note that $\Delta\xi(r)$ is related to the power spectrum $P(k)$ by
\begin{equation}
    \Delta\xi(r) = \int_0^\infty j_2(kr) {k^3 P(k)\over 2\pi^2} {dk\over k}.
\end{equation}
A similar two-dimensional statistic $\Delta\Sigma$ has been common in the weak lensing literature \citep{2001PhR...340..291B}.  As in that case, $\Delta\xi$ is unchanged by a constant offset in $\xi$ and therefore has zero support at $k=0$ and relative insensitivity to systematic errors at scales much larger than $r$.  As with the usual pair-counting methods, it is unaffected by shot noise.  However, it retains the localization of the acoustic scale information, here appearing as a dip rather than a peak.  It also avoids a zero crossing and is linear in $\xi$, so a logarithmic scaling shows the multiplicative scaling of large-scale bias.  A more complicated version of a statistic with this property was presented in \citet{2010ApJ...718.1224X}. 

Figure \ref{fig:corrfunc} shows the well-known progression of increasing bias with increasing halo mass \cite{1989MNRAS.237.1127C,1996MNRAS.282..347M} as well as the near scale-independence of bias.  At $z=1.1$, the mass bin above $10^{14}\hMsun$ is quite extreme, with a number density of only $1.6\times10^{-6}h^3{\rm\ Mpc}^{-3}$ and a bias exceeding 7!  These most massive halos show some large-scale scale-dependence in their clustering bias as well as an increased width of the acoustic peak.  These properties suggest care in interpreting the correlation function of high-mass clusters of galaxies.

\begin{figure}
    \centering
    \includegraphics[width=\columnwidth]{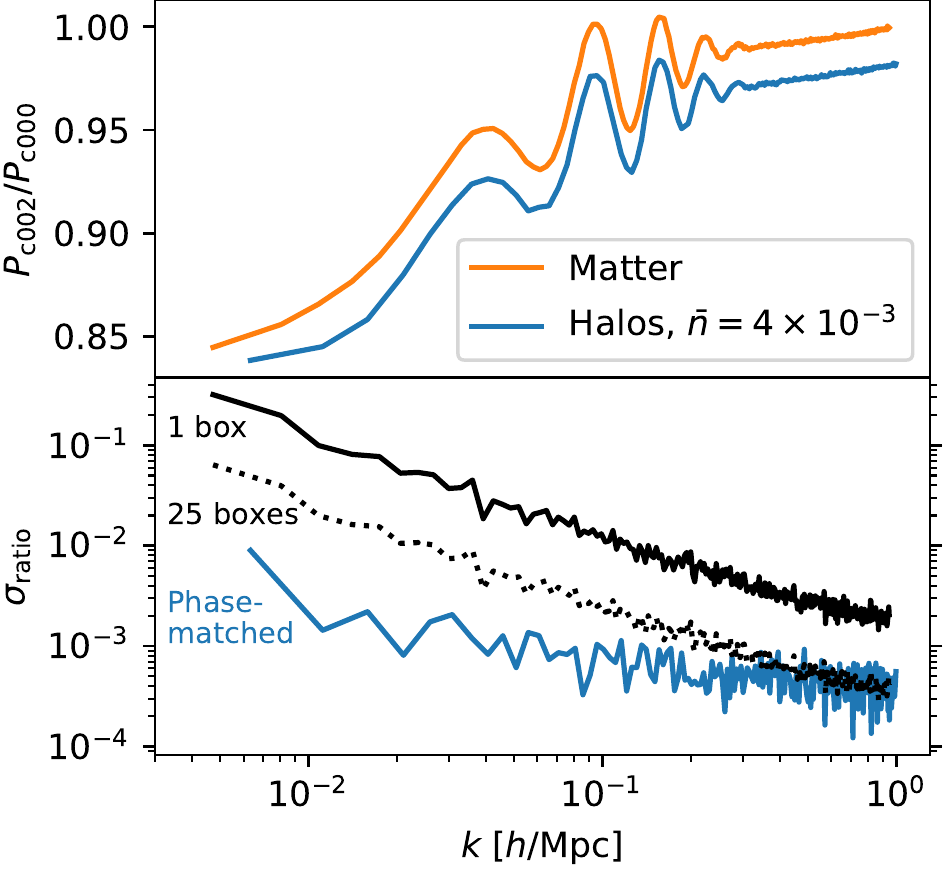}
    \caption{\textit{Top panel:} The ratio of $z=0.1$ power spectra from two phase-matched simulations that differ in cosmology---a derivative-like quantity.  The halo samples are abundance matched to $4\times10^{-3} h^3\mathrm{Mpc}^{-3}$.  \textit{Bottom panel:} Sample variances on the halo power spectrum from 1 box (black line), 25 boxes (black dotted line), and the variance on the ratio of 1 phase-matched pair (blue line).}
    \label{fig:phase-matched_derivatives}
\end{figure}

One of the design goals of \AbacusSummit{} is to provide the means to interpolate in a generous space of CDM cosmologies.  The use of phase-matched simulations is a key part of that, as we expect that this will suppress the sample variance in many applications to be comparable to or smaller than the variance that comes from the 25 \code{c000} base simulations.  If so, one can interpolate to new cosmologies while retaining the sample variance of a $200h^{-3}$~Gpc$^3$ volume.

In Figure~\ref{fig:phase-matched_derivatives}, we show an example comparing $\code{c000}$ to $\code{c002}$.  Both the matter power spectrum and halo power spectrum are shown, but we focus on the latter in the bottom panel as this is more representative of typical analysis.  The fractional sample variance on the power spectrum is shown for $8h^{-3}$~Gpc$^3$ (one box; black line) and $200h^{-3}$~Gpc$^3$ (25 boxes; black dotted line), and the sample variance on the power spectrum ratio is shown for one $8h^{-3}$~Gpc$^3$ phase-matched pair (blue line).  The phase-matched variance is smaller than the combined 25 boxes until about $k=1.0h\mathrm{Mpc}^{-1}$.  At $k=0.1h\mathrm{Mpc}^{-1}$, the phase-matching is equivalent to more than 50 boxes.  All measurements use $\Delta k=0.005$ power spectrum binning.

\subsection{Accuracy}

\begin{figure}[tb]
    \centering
    \includegraphics[width=\columnwidth]{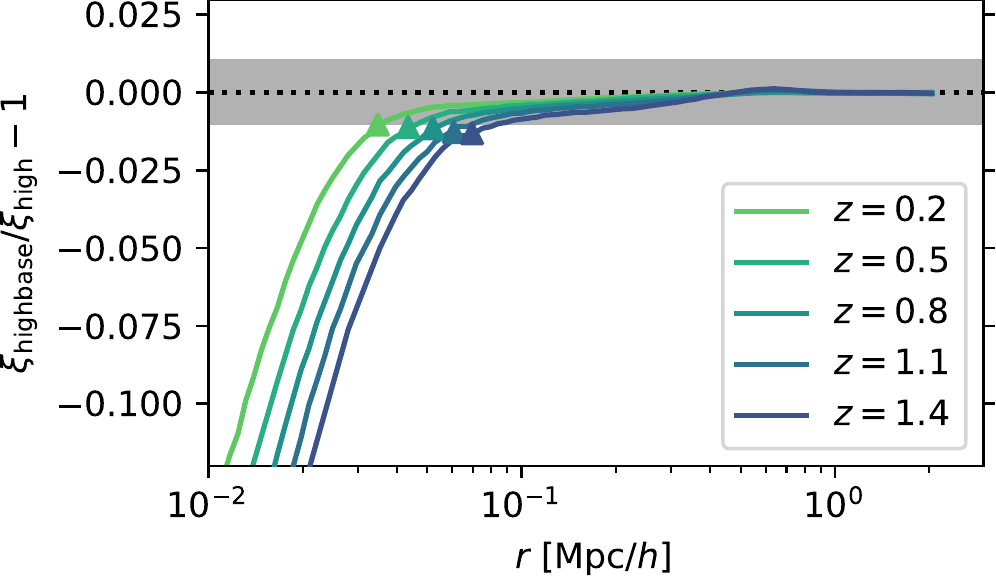}
    \caption{The small-scale matter correlation function of the \code{highbase} simulation (base mass resolution) compared with the \code{high} simulation ($6\times$ better mass resolution).  The triangles indicate the $4\times$ the comoving softening length of the \code{highbase} sim at that epoch (recalling that our softening is fixed in proper coordinates).}
    \label{fig:small_scale_2pcf}
\end{figure}

A major opportunity with the \Abacus{} code is its accuracy.  The near field is solved with explicit $O(N^2)$ summation, with no approximation from trees, out to a distance of $2\Mpc$.  Meanwhile, the multipole method generates very accurate far field forces, with modest errors from cells two units away and rapidly reaching machine precision beyond.  By comparing to control simulations in double-precision and with higher multipole order, we measure typical relative force errors of $10^{-5}$, far more accurate than codes that use particle-mesh far fields.

\AbacusSummit{} uses initial conditions with 2nd-order Lagrangian perturbation theory, which is known to provide substantial improvements over the 1st order theory (Zel'dovich approximation) for rare density peaks \citep{2006MNRAS.373..369C,2013MNRAS.431.1866R,2014NewA...30...79L}.  Moreover, we use the method of \citet{GarrisonEtal2016} to correct for the systematic under-evolution of continuum linear perturbation theory in a displaced discrete lattice.  The correction persists through intermediate redshifts, yielding a few-percent boost in the power near the Nyquist wavenumber of the particle lattice, matching what higher-resolution simulations yield.

The high speed of \Abacus{} coupled with the modest force-softening of our application allows us to proceed simply with global time stepping, avoiding the potential errors of individualized time steps.  \citet{GarrisonEtal2018} estimates that we have reached 1\% time-step convergence of the small-scale correlation function for our final answer.  

To further leverage this opportunity, we use a spline softening law that returns to the exact $1/r^2$ force at finite radius, unlike the commonly used Plummer softening based on the potential $1/\sqrt{r^2+\epsilon^2}$, which converges only quadratically to the correct large-scale force law.  This spline softening is moderately more expensive to compute, but the large hardware speed of the GPUs makes this an advantageous choice.
\citet{GarrisonEtal2018} shows the difference between these two softening choices, with Plummer softening showing percent-level modifications of the $z=0$ clustering even at $15\epsilon$.

As with any $N$-body simulation, there is a limit on spatial resolution due to the mass resolution of the particles and the force softening.  We have explored this using scale-free simulations \citep{2021MNRAS.501.5051J,2021MNRAS.501.5064L,Garrison+2021}, where one evolves a simulation from a power-law initial power spectrum.  These simulations show that a small force softening is not sufficient to reach a converged result for the small-scale correlation function; one is also limited by mass resolution, with a dependence that scales as $a^{-1/2}$.  Our softening length is chosen to capture the available convergence given the particle resolution, as shown in \citet{Garrison+2021}. 

To further demonstrate this, we here compare the small-scale matter correlation function between our highest mass resolution and base mass resolution, using the phase-matched pair \code{highbase\_c000\_ph100} to \code{high\_c000\_ph100}, with particle mass $2\times10^9\hMsun$ and $3.5\times10^8\hMsun$, respectively.
Figure \ref{fig:small_scale_2pcf} shows the ratio of the correlation functions at five redshifts.  We see that the two match to within 1\% to approximately 35 to $90\,h^{-1}\mathrm{kpc}$ between redshifts $0.2$ and $1.4$, decreasing with later epoch.  The point at which the difference reaches 1\% is approximately given by $4\epsilon(a)$, where $\epsilon(a)$ is the comoving softening length, recalling our softening is fixed in proper coordinates to $7.2\,h^{-1}\mathrm{kpc}$.  This scaling appears to differ from the $a^{-1/2}$ scaling in \citet{Garrison+2021}, although the mild change in scale factor in this plot makes the exact power-law ambiguous.  Two contributing reasons may be a) we have chosen a softening length that is comparable to the limit expected from the mass resolution, and b) we do not integrate with an arbitrarily short timestep.
For the latter, it is instructive to compute that timesteps of $10h^{-1}$~Myr imply that for a circular orbit in a high-mass cluster with circular velocity 1000 km/s would have $6\pi\approx20$ leapfrog steps per orbit at a radius of $30h^{-1}$~proper kpc (about 4 times the softening length).  Leapfrog integrations of two-body orbits with this number of steps do show reasonable conservation of radius, but develop a lag of about 10 degrees in phase per orbit.
The number of steps scales as radius divided by circular velocity, and integration accuracy scales as the  inverse square of the number of steps.  Given the sensitivity of the small-scale correlation function to high-mass halos, it is not surprising that we start to see some mild variations at a spatial scale where the most massive systems require orbital times shorter than a few dozen timesteps.


\begin{figure}
    \centering
    \includegraphics[width=\columnwidth]{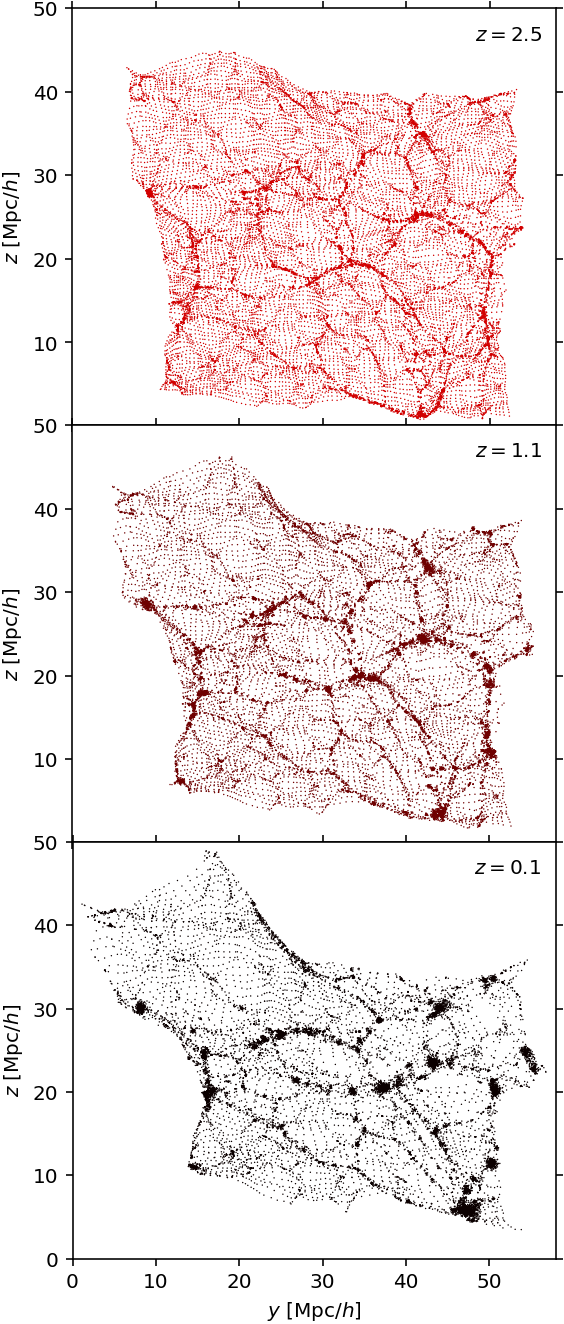}
    \caption{A single cutout of a Lagrangian plane---a square from a particle plane selected in the initial conditions---at redshifts $z=2.5$, $z=1.1$, and $z=0.1$ from the \code{highbase} simulation.  The ``memory'' of the initial lattice configuration persists to low redshift in low-density regions.}
    \label{fig:lattice_memory}
\end{figure}

The $z=1.4$ correlation function also exhibits an excess of about $0.1\%$ at $0.6\hMpc$, an effect attributed to the ``memory'' of the initial particle lattice.  
This is illustrated in Figure \ref{fig:lattice_memory}, which shows a single square of $150^2$ particles selected from a single Lagrangian sheet (a single particle plane in the initial conditions) at redshifts $z=2.5$ and $z=0.1$.  The lattice is seen quite strongly at the earlier time, as most particles have not yet fallen into halos or filaments and the deformations of the Lagrangian sheet are weak.  At low redshift, most particles have fallen into dense structures, but the voids become even sparser, and the remnants of the initial condition lattice are visible. 

We stress that this is endemic to any cosmological $N$-body simulation based on discrete particles; users must always take note that the regions that have not yet suffered any collapse will remember their initial configuration.  Randomized initial configurations like a glass make the effect less visibly obvious, but it would still be present.  Indeed, there are opportunities in uncollapsed regions to use the coldness of the initial sheet in phase space to estimate density and perform interpolations \citep[e.g.,][]{2012MNRAS.427...61A,2012PhRvD..85h3005S}; we note that the particle ID numbers in the full time-slice outputs could be helpful in such work.

As a final caveat, we note that while the $N$-body problem is crisply posed and we believe that \Abacus{} offers very accurate solutions thereof, the identification of halos and the associated interpretations of galaxy locations are always subject to choices.  There will surely be differences between CompaSO halos and those of other methods, and it is possible that these differences can affect mock galaxy catalogs in ways that matter for cosmological analysis.  Users of any $N$-body simulation must consider this.

\section{Conclusion}\label{sec:conclusion}

The \AbacusSummit{} simulation suite constitutes a uniquely large and diverse data-set designed to enable a broad range of science applications in preparation for the next generation of large-scale structure cosmology. Simulating roughly 60 trillion particles, the suite constitutes the largest $N$-body data-set produced to date. \AbacusSummit has been designed to provide a large-volume, high-accuracy simulation of the Planck2018 $\Lambda$CDM cosmology, as well as  a grid of 96 other cosmologies to allow for interpolation in parameter space and emulator construction. \AbacusSummit likewise includes 1883 small boxes to support covariance matrix estimation under periodic boundary conditions, and an array of simulations designed to facilitate code comparison studies, group finding and mass resolution studies, as well as comparisons to major flagship simulations from other codes.  \AbacusSummit simulations include a diverse set of data-products which have undergone extensive validation, including particle subsamples, halo catalogs, merger trees, kernel density estimates, light cones, and projected density maps of the light cones. \AbacusSummit was produced using the \Abacus code and is available publicly at a DOI link published on \url{https://abacussummit.readthedocs.io/}.

\acknowledgements\label{sec:Acknowledgements}

We thank Philip Pinto for his advice on the project and  Peter Behroozi, Shaun Cole, Salman Habib, Katrin Heitmann, Alina Kiessling, and Joachim Stadel for helpful conversations.

This work has been supported by NSF AST-1313285, DOE-SC0013718, and NASA ROSES grant 12-EUCLID12-0004.
DJE is supported in part as a Simons Foundation investigator. 
NAM was supported in part as a NSF Graduate Research Fellow.  
LHG is supported by the Center for Computational Astrophysics at the Flatiron Institute, which is supported by the Simons Foundation.  
SB is supported by the UK Research and Innovation (UKRI) Future Leaders Fellowship (grant number MR/V023381/1), and formerly by Harvard University through the ITC Fellowship.

This research used resources of the Oak Ridge Leadership Computing Facility, which is a DOE Office of Science User Facility supported under Contract DE-AC05-00OR22725. 
Computation of the merger trees used resources of the National Energy Research Scientific Computing Center (NERSC), a U.S. Department of Energy Office of Science User Facility located at Lawrence Berkeley National Laboratory, operated under Contract No. DE-AC02-05CH11231.
The \AbacusSummit simulations have been supported by OLCF projects AST135 and AST145, the latter through the Department of Energy ALCC program.

We would like to thank the OLCF and NERSC staff for their highly responsive and expert assistance, both scientific and administrative, during the course of this project.

\software{Astropy \citep{Astropy_2018},
            NumPy \citep{van_der_Walt+2011},
            SciPy \citep{Virtanen+2020},
            Numba \citep{Lam+2015},
            CUDA \citep{Nickolls+2008},
            Intel TBB \citep{Reinders_2007},
            matplotlib \citep{Hunter_2007},
            ASDF \citep{GREENFIELD2015240},
            Globus \citep{Foster_2011,Allen+2012},
            Corner.py \citep{corner},
            Corrfunc \citep{Sinha_Garrison_2019,Sinha_Garrison_2020},
}

\section*{Data Availability}
With this paper, we are placing the \AbacusSummit{} simulation suite into the public domain, subject to the academic citations described at \url{https://abacussummit.readthedocs.io/en/latest/citation.html}.

Data access is available through OLCF's Constellation portal.  The persistent DOI describing the data release is \href{https://www.doi.org/10.13139/OLCF/1811689}{10.13139/OLCF/1811689}.  Instructions for accessing the data are given at \url{https://abacussummit.readthedocs.io/en/latest/data-access.html}.

\bibliography{bibliography}
\bibliographystyle{aasjournal}

\end{document}